\documentclass[a4paper,12pt]{article}

\usepackage{amssymb,amsmath,empheq,epsfig,graphicx,cite}

\newcommand{\A}{\mathcal{A}}             % action
\newcommand{\Auno}{A_1}                  % parameter 9* a_2
\newcommand{\az}{\alpha_0}
\newcommand{\be}{\beta}
\newcommand{\bk}[1]{\langle #1 \rangle}  % bra-ket
\newcommand{\bs}{\boldsymbol}
\newcommand{\bra}[1]{\langle#1|}         % <   |
\newcommand{\cl}{\mathrm{cl}}            % classical
\newcommand{\dif}{\mathrm{d}}            % finite-dimensional differential
\newcommand{\Dif}[1]{[\mathcal{D} #1 ]}  % infinite-dimensional differential
\renewcommand{\div}{\mathrm{div}}
\newcommand{\el}{\mathrm{el}}           % elastic
\newcommand{\esp}[1]{\mathrm{e}^{#1}}    % esponential
\newcommand{\Hc}{\mathcal{H}}
\newcommand{\half}{{\textstyle\frac{1}{2}}}
\newcommand{\imp}{\Longrightarrow}
\newcommand{\Ket}[1]{|\{#1\}\rangle}     % |{ }>
\newcommand{\ket}[1]{|#1\rangle}         % |   >
\newcommand{\kt}{\boldsymbol{k}}
\newcommand{\N}{\mathbb{N}}              % natural
\newcommand{\nplus}{\tilde{n}}           % n+1/2
\newcommand{\ord}[1]{\mathcal{O}\left(#1\right)}
\newcommand{\quarter}{{\textstyle\frac{1}{4}}}
\newcommand{\reg}{\mathrm{reg}}
\newcommand{\rinf}{\rho_{\infty}}        % rho_2(inf)
\newcommand{\sign}{\mathrm{sign}}        % sign
\newcommand{\threehalf}{{\textstyle{\frac32}}}  % 3/2
\newcommand{\tz}{t_0}
\newcommand{\U}{\mathcal{U}}
\newcommand{\ui}{\mathrm{i}}             % imaginary unit
\newcommand{\xt}{\boldsymbol{x}}

\title{{ Unitarity Alternatives in the Reduced-action Model for Gravitational Collapse}}

\author{
   M.~Ciafaloni, D.~Colferai and G.~Falcioni\\[1ex]
   \sl Dipartimento di Fisica, Universit\`a di Firenze and\\
   INFN, Sezione di Firenze, 50019 Sesto Fiorentino, Italy
}
 
\date{}

\begin{document}
%@@@@@@@@@@@@@@@@@@@@@@@@@@@@@@@@@@@@@@@@@@@@@@@@@@@@@@@@@@@@@@@@@@@@@@@@@@@@@@@

\maketitle

\begin{abstract}
  Based on the ACV approach to transplanckian energies, the reduced-action model
  for the gravitational $S$-matrix predicts a critical impact parameter
  $b_c\sim R\equiv 2G\sqrt{s}$ such that $S$-matrix unitarity is satisfied in
  the perturbative region $b>b_c$, while it is exponentially suppressed with
  respect to $s$ in the region $b<b_c$ that we think corresponds to
  gravitational collapse. Here we definitely confirm this statement by a
  detailed analysis of both the critical region $b\simeq b_c$ and of further
  possible contributions due to quantum transitions for $b<b_c$. We point out,
  however, that the subcritical unitarity suppression is basically due to the
  boundary condition which insures that the solutions of the model be
  ultraviolet-safe. As an alternative, relaxing such condition leads to
  solutions which carry short-distance singularities presumably regularized by
  the string. We suggest that through such solutions --- depending on the
  detailed dynamics at the string scale --- the lost probability may be
  recovered.
\end{abstract}

\vskip 1cm

%\begin{minipage}{0.9\textwidth}
%\begin{flushright}
%  Draft $ $Revision: 143 $ $ \\
%  $ $Date: 2011-06-28 12:36:03 +0200 (Tue, 28 Jun 2011) $ $
%\end{flushright}
%\end{minipage}
\vskip 1cm

%%%%%%%%%%%%%%%%%%%%%%%%%%%%%%%%%%%%%%%%%%%%%%%%%%%%%%%%%%%%%%%%%%%%%%%%%%%%%%%%%
\section{Introduction\label{s:intro}}
%%%%%%%%%%%%%%%%%%%%%%%%%%%%%%%%%%%%%%%%%%%%%%%%%%%%%%%%%%%%%%%%%%%%%%%%%%%%%%%%%

Interest in the gravitational $S$-matrix at transplanckian
energies~\cite{ACV88,GrMe87,ACV90,ACV93} has revived in the past
few-years~\cite{ACV07,CC08,CC09}, when explicit solutions of the so-called
reduced-action model~\cite{ACV93} have been found~\cite{ACV07}. The model is a
much simplified version of the ACV eikonal approach~\cite{ACV88,ACV90} to
transplanckian scattering in string-gravity, and is valid in the regime in which
the gravitational radius $R\equiv2G\sqrt{s}$ is much larger than the string
length $\lambda_s\equiv\sqrt{\alpha'\hbar}$, so that string-effects are supposed
to be small.

The reduced-action model~(sec.~\ref{s:raa}) was derived by justifying the
eikonal form of the $S$-matrix at impact parameter $b$ on the basis of string
dynamics and by then calculating the eikonal itself (of order
$\sim\frac{Gs}{\hbar}\gg1$) in the form of a 2-dimensional action, whose power
series in $\frac{R^2}{b^2}$ corresponds to an infinite sum of proper irreducible
diagrams (the ``multi-H'' diagrams~\cite{ACV90,ACV93}), evaluated in the
high-energy limit. The model admits a quantum generalization~\cite{CC08} of the
$S$-matrix in the form of a path-integral --- with definite boundary conditions
--- of the reduced-action exponential itself.

The main feature of the model and of its boundary conditions is the existence of
a critical impact parameter $b_c\sim R$ such that, for $b>b_c$ the $S$-matrix
matches the perturbative series and is unitary, while for $b<b_c$ the field
solutions are complex-valued and the elastic $S$-matrix is suppressed
exponentially. The suppression exponent is of order
$\frac{Gs}{\hbar}\sim\frac{R^2}{\lambda_P^2}$ ($\lambda_P$ being the Planck
length) or, if we wish, of the same order as the entropy of a black-hole of
radius $R$. From various arguments we believe that in the region in which
$b<b_c$ (that is, $b$ is smaller than the gravitational radius), a
classical gravitational collapse is taking place.

The model, in its simplest axisymmetric form, is formulated in terms of only one
effective field $\rho(r^2)$ depending on the transverse radius squared
$r^2\equiv\xt^2$ and is defined by
$\rho(r^2)\equiv r^2\left(1-(2\pi R)^2\frac{\dif}{\dif r^2}\phi\right)$,
where $h=\nabla^2\phi$ determines the transverse gravitational field and the
corresponding metric, which is of shock-wave type. A key role
in the derivation of the above features is played by the boundary condition
$\rho(0)=0$ which avoids a possible singularity of $h$ at $r^2=0$ and is the
main cause of the suppression of the $S$-matrix for $b<b_c$. In fact, the
complex solutions of the semiclassical approach for $b<b_c$ were interpreted at
quantum level~\cite{CC08} as due to a tunnel effect, required in order to reach
$\rho=0$ at $r^2=0$, across a potential barrier occurring in the lagrangian.

At this point the question was (and is): do we reach $S$-matrix unitarity for
$b<b_c$ by summing over inelastic processes? do we recover full
information~\cite{Hawking} from the scattering experiment in the collapse
region? It was already found in~\cite{CC09}, by semiclassical methods, that this
is not the case, and that the unitarity defect persists when all inelastic
channels are included, although in a way dependent on the rapidity phase space
parameter $Y$. This result is puzzling because one would like to know where does
the probability go if the model is complete or --- if it is not --- how to
complete it.

The purpose of the present paper is both to look at possible flaws in the result
just quoted and to suggest a tentative answer to the ensuing question. We
exclude flaws in two ways. In sec.~\ref{s:sud} we perform a detailed analysis of
the solutions of the semiclassical unitarity equations, we choose the stable
ones and we investigate the unitarity behaviour by a perturbative method around
the critical point $b\simeq b_c$ and by numerical methods elsewhere. There are
no surprises: the results of~\cite{CC09} are fully checked, we only gain some
better understanding of the unitarity defect around the critical point.

As a second attempt, we look for a quantum treatment of inelastic $S$-matrix
elements~(sec.~\ref{s:sqt}). Since the quantum version of the reduced action
model features a quantum mechanical hamiltonian with a Coulomb potential in
$\rho$-space, we eventually evaluate quantum transitions from the basic
tunneling wave function to other states of the system. Of particular interest
are the bound states in the strong-coupling region $\rho\leq0$ where the
potential is attractive, because they could correspond to collapsed
matter. Unfortunately, all relevant matrix elements carry the same exponential
suppression as the tunneling amplitude itself, and the unitarity defect
survives.

As a final point, in sec.~\ref{s:rhof} we test the boundary conditions of the
model, by letting $\rho(0)$ fluctuate away from zero, with the weight assigned
to it by the reduced action itself. We find that, while the elastic $S$-matrix
element is stable --- that is, dominated by very small $\rho(0)$ --- large
inelastic contributions may come from the short-distance region, where however
the model is inadequate and string effects are expected to play an important
role. We argue on this basis that in the direction of such ultraviolet-sensitive
solutions the model is incomplete and that, by completing it with the proper
string dynamics we may discover where the probability goes.

%%%%%%%%%%%%%%%%%%%%%%%%%%%%%%%%%%%%%%%%%%%%%%%%%%%%%%%%%%%%%%%%%%%%%%%%%%%%%%%%
\section{The reduced-action model for gravitational $\bs{S}$-matrix\label{s:raa}}
%%%%%%%%%%%%%%%%%%%%%%%%%%%%%%%%%%%%%%%%%%%%%%%%%%%%%%%%%%%%%%%%%%%%%%%%%%%%%%%%

We provide here a brief account of the model being considered --- as develope in
refs.~\cite{ACV07,CC08} --- for both completeness sake and in order to emphasize
some points which are useful in the following.

%===============================================================================
\subsection{The semiclassical ACV results\label{s:sacvr}}
%===============================================================================

The simplified ACV approach~\cite{ACV07} to transplanckian scattering in the
regime $R \equiv 2G\sqrt{s} \gg \lambda_s$ is based on two main points.
Firstly, the gravitational field $g_{\mu\nu}=\eta_{\mu\nu}+h_{\mu\nu}$
associated to the high-energy scattering of light particles, reduces to a
shock-wave configuration of the form
\begin{subequations}\label{metrica}
  \begin{align}
    h_{--}\big|_{x^+=0} &= (2\pi R)a(\xt)\delta(x^-) \;, \qquad
    h_{++}\big|_{x^-=0} = (2\pi R)\bar{a}(\xt)\delta(x^+)
    \label{hlong} \\
    h_{ij} &= (\pi R)^2\Theta(x^+ x^-)
    \left(\delta_{ij}-\frac{\partial_i\partial_j}{\nabla^2}\right) h(\xt) \;,
    \label{hij}
  \end{align}
\end{subequations}
where $a$, $\bar{a}$ are longitudinal profile functions, and
$h(\xt)\equiv\nabla^2\phi$ is a scalar field describing one emitted-graviton
polarization (the other, related to soft graviton radiation, is negligible in an
axisymmetric configuration).

Secondly, the high-energy dynamics itself is summarized in the $h$-field
emission-current $\Hc(\xt)$ generated by the external sources coupled to the
longitudinal fields $a$ and $\bar{a}$. Such a vertex has been calculated long
ago~\cite{Li82,ABC89} and takes the form
\begin{equation}\label{vertex}
  -\nabla^2 \Hc \equiv \nabla^2 a \nabla^2\bar{a}-\nabla_i\nabla_j a
  \nabla_i\nabla_j\bar{a} \;,
\end{equation}
which is the basis for the gravitational effective action~\cite{Li91,KiSz95,Ve93}
from which the shock-wave solution~(\ref{metrica}) emerges~\cite{ACV93}. It is
directly coupled to the field $h$ and, indirectly, to the external sources $s$
and $\bar{s}$ in the reduced 2-dimensional action
\begin{equation}\label{2dimAction}
  \frac{\A}{2\pi Gs} = \int\dif^2\xt\left[ a\bar{s}+\bar{a}s-\frac12\nabla a
    \nabla\bar{a}+\frac{(\pi R)^2}{2}\left(-(\nabla^2\phi)^2-2\nabla\phi\cdot
\nabla\Hc\right)\right]
\end{equation}
which is the basic ingredient of the ACV simplified treatment. Note that here
the gravitational radius $R$ plays the role of (dimensionful) coupling constant
and that --- because of the higher derivatives of $\phi$ involved ---
non-renormalizable UV divergences may occur in general.

\begin{figure}[ht!]
  \centering
  \includegraphics[height=0.15\textwidth]{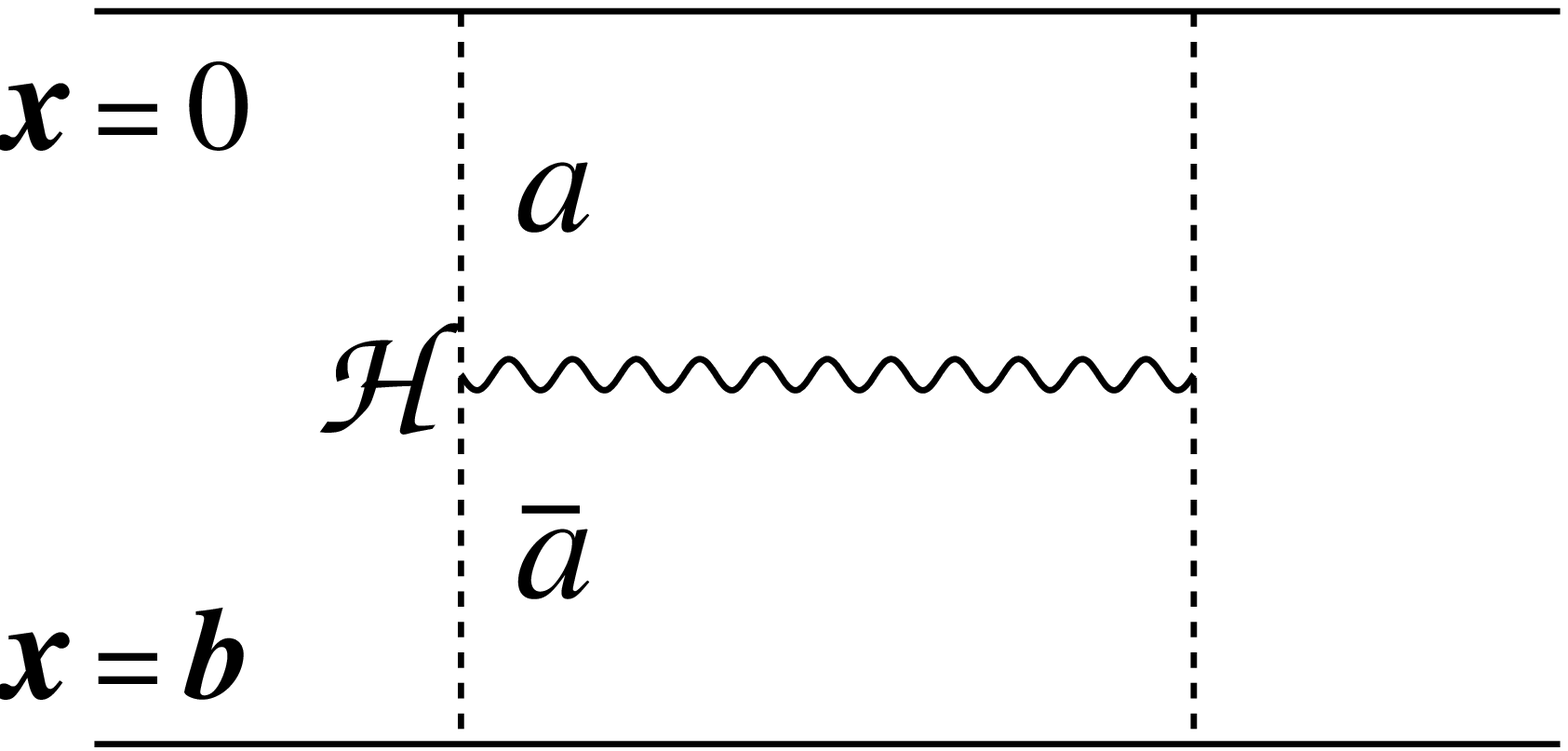}
  \hspace{1em}\raisebox{0.07\textwidth}{+} \hspace{1em}
  \includegraphics[height=0.15\textwidth]{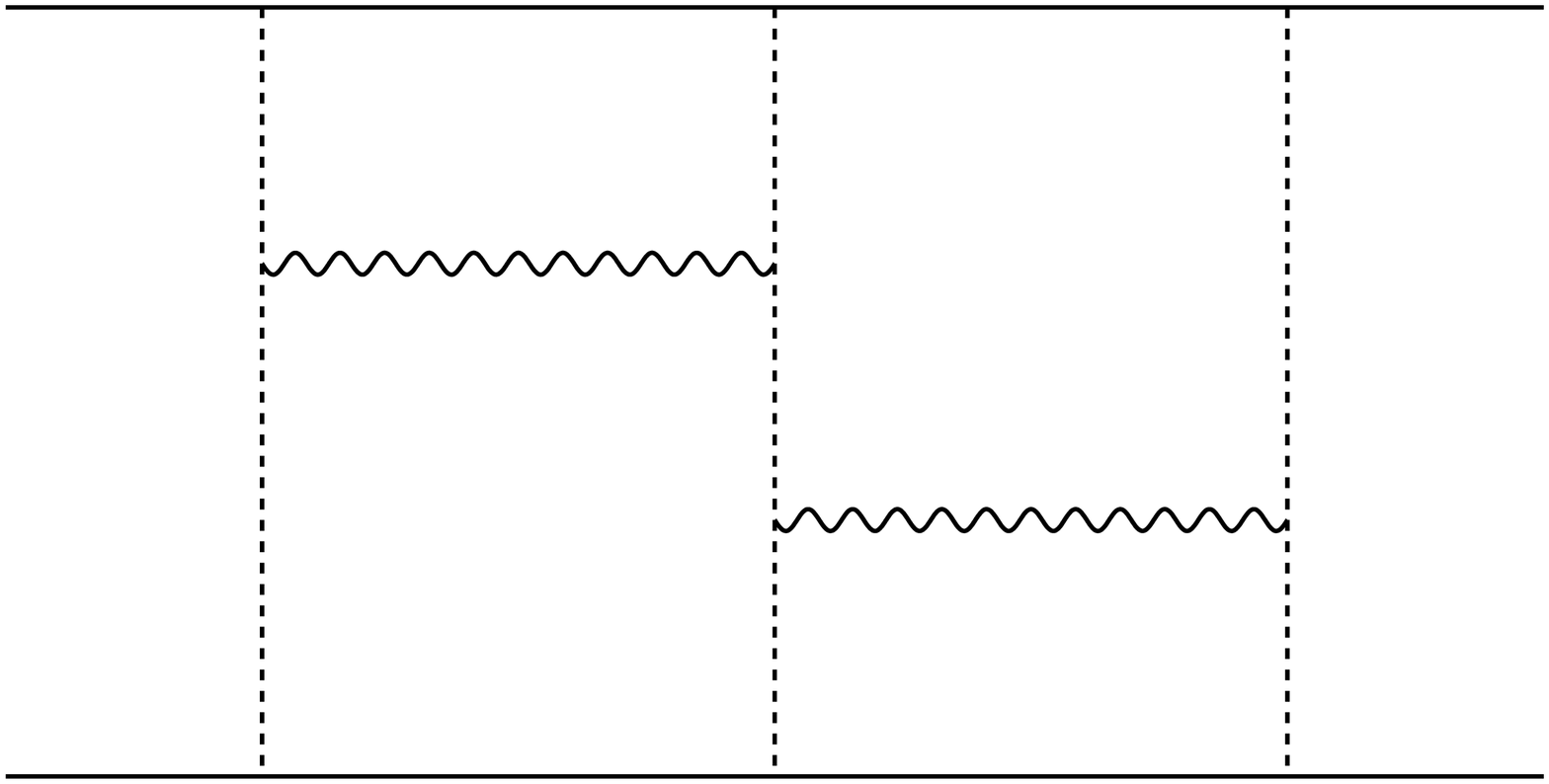}
  \hspace{1em} \raisebox{0.07\textwidth}{+ \dots}
  \caption{\it Diagrammatic series of H and multi-H diagrams.}
  \label{f:multiH}
\end{figure}

The equations of motion (EOM) induced by (\ref{2dimAction}) provide, with proper
boundary conditions, some well-defined effective metric fields $a$ and $h$ which
are, hopefully, UV-safe. The ``on-shell'' action $\A (b,s)$, evaluated on such
fields, provides directly the elastic $S$-matrix
\begin{equation}\label{elSmatrix}
  S_\el = \exp\left( \frac{\ui}{\hbar} \A(b,s) \right).
\end{equation}
Then, it can be shown~\cite{ACV93,ACV07} that the reduced-action above (where
now $R/b$ plays the role of effective coupling constant) resums the so-called
multi-H diagrams (fig.~\ref{f:multiH}), contributing a series of corrections
$\sim (R^2/b^2)^n$ to the leading eikonal, as well as their resummation for
$R/b=\ord{1}$.

Furthermore, the $S$-matrix (\ref{elSmatrix}) can be extended to inelastic
processes on the basis of the same emitted-graviton field $h(\xt)$.  In the
eikonal formulation the inelastic $S$-matrix is approximately%
\footnote{The coherent state describes uncorrelated emission (apart from
  momentum conservation~\cite{CVprep}). However, the eikonal approach based on
  eq.~(\ref{2dimAction}) also predicts~\cite{ACV07} correlated particle
  emission, which is suppressed by a power of $(Gs/\hbar) Y$ relative to the
  uncorrelated one, and is not considered here.\label{fn:cs}}
described by the coherent state operator
\begin{align}
 &S = \exp\left( \frac{\ui}{\hbar} \A(b,s) \right)
 \exp\left( \ui 2\pi R \sqrt{\alpha}
 \int\dif^2 \xt \; h(\xt) \Omega(\xt) \right)
 \;, \qquad \alpha \equiv \frac{Gs}{\hbar}
 \label{Scsh} \\
 &\Omega(\xt) \equiv \int\frac{\dif^2\kt\,\dif k_3}{2\pi\sqrt{k_0}}
 \left[ a(\kt,k_3) \esp{\ui\kt\cdot\xt} + h. c. \right]
 \equiv A(\xt)+A^\dagger(\xt)\;,
  \label{Omega}\\
 &[ A(\xt), A^\dagger(\xt')] = Y \delta(\xt - \xt')
  \nonumber
\end{align}
where the operator $\Omega(\xt)$ incorporates both emission and absorption of
the $h$-fields and $Y$ parameterizes the rapidity phase space which is
effectively allowed for the production of light particles (e.g.~gravitons).

For a given value of the ``gravitational coupling'' $\alpha\equiv Gs/\hbar$
the parameter $Y$ is possibly large for
large impact parameters $b \gg \sqrt{G\hbar}$, because the effective transverse
mass of the light particles is expected to be of order $\hbar/b$, i.e., much
smaller than the Planck mass, thus yielding roughly
$Y \sim \log(s b^2/\hbar^2) \gg 1$. On the other hand, we should notice that
dynamical arguments based on energy conservation~\cite{CVprep} and on absorptive
corrections of eikonal type, consistent with the AGK cutting rules~\cite{AGK73},
tend to suppress the fragmentation region in a $b$-dependent way, so as to
constrain $Y$ to be $\ord{1}$ for impact parameters in the classical collapse
region $b=\ord{R}$.  However, such arguments do not take into account possible
dynamical correlations coming from multi-H diagrams, as mentioned in
footnote~\ref{fn:cs}. It is fair to state that a full dynamical understanding of
the $Y$ parameter is not available yet, and for this reason we shall discuss
what happens for any values of $Y$.

In the case of axisymmetric solutions, where $a=a(r^2)$, $\bar{a}=\bar{a}(r^2)$,
$\phi=\phi(r^2)$ it is straightforward to see, by using eq.~(\ref{vertex}), that
$\dot{\Hc}(r^2)\equiv(\dif/\dif r^2)\Hc(r^2)=-2\dot{a}\dot{\bar{a}}$ becomes
proportional to the $a,\bar{a}$ kinetic term. Therefore, the
action~(\ref{2dimAction}) can be rewritten in the more compact one-dimensional
form
\begin{equation}\label{1dimAction}
 \frac{\A}{2\pi^2 Gs}=\int\dif r^2\left[ a(r^2)\bar{s}(r^2) + \bar{a}(r^2)s(r^2)
   -2\rho\dot{\bar{a}}\dot{a} - \frac{2}{(2\pi R)^2}(1-\dot\rho)^2\right]
 \;, \qquad \dot{a} \equiv\frac{\dif a}{\dif r^2} \;,
\end{equation}
where we have introduced the auxiliary field $\rho(r^2)$
\begin{equation}\label{rho}
  \rho=r^2\big(1-(2\pi R)^2\dot\phi\big) \;, \qquad
  h=4\dot{(r^2\;\dot{\!\!\!\phi})\!}=\frac1{(\pi R)^2}(1-\dot\rho)
\end{equation}
which incorporates the $\phi$-dependent interaction. The external sources
$s(r^2)$, $\bar{s}(r^2)$ are assumed to be axisymmetric also, and are able to
approximately describe the particle-particle case by setting
$\pi s(r^2)=\delta(r^2)$, $\pi\bar{s}(r^2)=\delta(r^2-b^2)$, where the azimuthal
averaging procedure of ACV is assumed.%
\footnote{The most direct interpretation of this configuration is the scattering
  of a particle off a ring-shaped null matter distribution, which is
  approximately equivalent to the particle-particle case by azimuthal
  averaging~\cite{ACV07}.}

The equations of motion, specialized to the case of particles at impact
parameter $b$ have the form
\begin{align}
  \dot{a} &= -\frac1{2\pi\rho} \;, \qquad
  \dot{\bar{a}} = -\frac1{2\pi\rho}\Theta(r^2-b^2) \;, \label{eoma}\\
  \ddot{\rho} &= \frac1{2\rho^2}\Theta(r^2-b^2)
  % \;, \qquad \dot{\rho}^2+\frac1{\rho} = 1
  \qquad (r > b) \label{eomrho}
\end{align}
and show a ``Coulomb'' potential in $\rho$-space, which is repulsive for
$\rho>0$, acts for $r>b$ and plays an important role in the tunneling
phenomenon. By replacing the EOM~(\ref{eoma}) into eq.~(\ref{1dimAction}), the
reduced action can be expressed in terms of the $\rho$ field only, and takes the
simple form
\begin{equation}\label{rhoAction}
  \A = -Gs\int\dif r^2
  \left(\frac1{R^2}(1-\dot\rho)^2-\frac1{\rho}\Theta(r^2-b^2)\right)
  \equiv -\int_0^\infty \dif r^2 \; L(\rho,\dot\rho,r^2) \;,
\end{equation}
which is the one we shall consider at quantum level in the following.

The effective metric~(\ref{metrica}) generated by the axisymmetric fields $\rho$, $a$ and
$\bar{a}$ is calculated~\cite{ACV07} on the basis of the complete form of the
shock-wave~(\ref{metrica}) and is given by
\begin{align}
 \dif s^2 &= -\dif x^+\dif x^- 
 \left[1-\half \Theta(x^{+}x^{-})(1-\dot\rho)\right]
 \nonumber \\
 &+(\dif x^+)^2 \delta(x^{+}) \left[
   2\pi R \bar{a}(r^2) -\quarter(1-\dot\rho) |x^{-}|
 \right]
 \nonumber \\
 &+(\dif x^-)^2 \delta(x^{-}) \left[
   2\pi R a(r^2) -\quarter(1-\dot\rho) |x^{+}|
 \right]
 \nonumber \\
 &+ \dif r^2\left[1+2(\pi R)^2\Theta(x^{+}x^{-})\dot{\phi} \right]
 +\dif\theta^2\,r^2 \left[1+2(\pi R)^2(\dot{\phi}+r^2\ddot{\phi})\right]
 \label{metr}
\end{align}
This metric is dynamically generated and may be regular or singular at short
distances, depending on the behaviour of the field solutions themselves.

%===============================================================================
\subsection{Boundary conditions and critical impact parameter\label{s:bccip}}
%===============================================================================

Two boundary conditions are necessary to solve the equation of motion~(\ref{eomrho}):
first of all we set $\dot{\rho}(\infty)=1$ in order to have a gravitational
field $h\sim 1-\dot{\rho}$ vanishing at large distances.  The second boundary
condition is $\rho(0)=0$ and it is necessary in order to obtain UV-safe
solutions at short distances. Indeed, by the definition of $\rho$ --- which
embodies an $r^2$ factor in eq.~(\ref{rho}) ---, the fields
$\phi$ and $h$ are singular if $\rho(0)\neq 0$. More precisely
$\dot\phi \simeq -\rho(0)/r^2$ is quadratically divergent, and this implies that
the outgoing flux of $\nabla\phi$ at the origin is $\simeq -\rho(0)$
and therefore $h \equiv \nabla^2\phi \simeq -\rho(0)\delta(r^2)$ is singular too. Thus, solutions
with $\rho(0) \neq 0$ possess UV singularities, and this implies in turn an UV
divergent action and a singular metric. In particular, the singularity of
$\dot{\phi}$ is present in the metric coefficient $h_{rr}$, which changes sign
also, while a $\delta(r^2)$ singularity of difficult interpretation occurs in
the rest of the metric (\ref{metr}). For these reasons we set $\rho(0) = 0$ so
as to obtain a finite action and a regular metric.

The solution of the equation of motion satisfying the above boundary conditions
is
\begin{equation}\label{soluz}
  \rho=
  \left\{
  \begin{array}{l}
    t_{b}r^2 \qquad\qquad\qquad\;\; (r^2<b^2)\\[0.5ex]
    R^2\cosh^2\chi(r^2) \qquad (r^2 \geq b^2) \\[0.5ex]
    r^2-b^2=R^2(\chi+\sinh\chi \cosh\chi-\chi_{b}-\sinh\chi_{b}\cosh\chi_{b})
  \end{array} 
  \right.
\end{equation}
with $\tanh{\chi_{b}}=t_{b}=\dot{\rho}(b^2)$ and $\chi_b\equiv\chi(b^2)$. The parameter $t_{b}$ is
determined by requiring regular matching of the solution at $r^2=b^2$, that is
\begin{equation}\label{matching}
   t_{b}(1-t_{b}^2)=\frac{R^2}{b^2}
\end{equation}
This equation acquires the meaning of criticality equation. In fact it can be
solved%
\footnote{We consider only solutions with $t_{b}>0$, otherwise the condition
  $\dot{\rho}(\infty)=1$ would not be satisfied.}
only if the impact parameter $b$ exceeds a critical value $b_c$ given by
$b_c^2=\frac{3\sqrt{3}}{2}R^2$. For $b>b_c$, one of the two real solutions
matches the perturbative result~\cite{ACV90,ACV07}
\begin{equation}\label{Abs}
  \A(b,s) = Gs\left(\log\frac{L^2}{b^2}+\frac{R^2}{2b^2}+\cdots\right) \;,
\end{equation}
where $L$ is an infrared cutoff parameterizing the infinite ``Newtonian'' phase.
If instead $b<b_c$, the criticality equation has two complex conjugate
solutions, providing an imaginary part of $\rho$ in eq.~(\ref{soluz}) that we
can't interpret at purely classical level. Anyway we can try to use these
complex solutions and substitute them in the on shell action, which is easily
calculated to be
\begin{equation}\label{A}
 \A = Gs\left\{2[\chi(L^2)-\chi_{b}]-\frac{1-t_b}{t_b}\right\}
\end{equation}
where $\chi(L^2) \simeq \log(L/R)$.

For $b<b_c$ this expression becomes complex-valued, providing a non unitary
$S=\esp{\ui\A}$. It can be shown by stability arguments~\cite{ACV07} that the
physical solution of (\ref{matching}), when $b<b_c$, is the one with negative
imaginary part of $t_{b}$, giving rise to a suppression of the $S$-matrix in the
elastic channel. For example, for $b=0$, we obtain $\chi_{b=0} = -\ui\pi/2$ and
the exponential suppression
\begin{equation}\label{b=0}
  S\sim \esp{-\pi G s} \;.
\end{equation}
The suppression exponent $Gs\sim R^2/\lambda_P^2$ is of the same order as the
entropy of a black-hole of radius $R$, but occurs here because of the
complex-valued $\rho(r^2)$ when the impact parameter is smaller than the
critical value.

We conclude this subsection by noting that real-valued solutions to
eq.~(\ref{eomrho}) and satisfying $\dot\rho(\infty)=1$ are always of the
form~(\ref{soluz}) and have necessarily $\rho(0)>0$ whenever $b<b_c$. The
minimum (and closest to 0) value of $\rho(0)$ is found for a particular value of
the initial slope $\dot\rho(0)=t_m$ determined by
\begin{equation}\label{tm}
  \frac{b^2}{2R^2} = \cosh^3(\chi_m)\sinh(\chi_m)=\frac{t_m}{(1-t_m^2)^2} \;.
\end{equation}
The corresponding solution provides, so to say, the real solution with smallest
``distance'' to the complex solution having $\rho(0)=0$.

%===============================================================================
\subsection{The quantized $\bs{S}$-matrix and tunnel effect\label{s:qsm}}
%===============================================================================

The suppression of the $S$-matrix for $b < b_c$ can be interpreted by
generalizing the scattering matrix to a quantum level. In the CC
proposal~\cite{CC08} this is achieved by summing the reduced-action exponential
over every path $\rho(r^2)$ satisfying the boundary conditions $\rho(0)=0$ and
$\dot{\rho}(\infty)=1$
\begin{equation}\label{d:Sq}
 S(b, s)= \int_{\rho(0)=0}^{\dot{\rho}(\infty)=1}
 \Dif{\rho(r^2)} \; \esp{-\ui\int L(\rho,\dot{\rho},r^2) \dif r^2}\,
 \esp{\frac{2\ui\sqrt{\alpha}}{\pi R}\int
 [1-\dot{\rho}]\Omega(\xt) \; \dif^2\xt} \;,
\end{equation}
where we have included the coherent state operator $\Omega$ of eq.~(\ref{Omega})
in order to describe inelastic processes. The $S$-matrix in the elastic channel
is obtained by evaluating the vacuum expectation value and we obtain
\begin{align}
 S_\el &= \int_{\rho(0)=0}^{\dot{\rho}(\infty)=1}
 \Dif\rho\; \esp{-\ui\int L_{y}(r^2)\; \dif r^2}
 \label{sel} \\
 L_y &= \frac{1}{4G}\left[(1-\ui y)(1-\dot{\rho})^2
   -\frac{R^2}{\rho}\Theta(r^2-b^2)\right] \;,
  \qquad y \equiv \frac{2Y}{\pi} \;.
 \label{Ly}
\end{align}
The parameter $y$ is related to the rapidity phase space allowed
for the emitted gravitons, and provides a related absorption of the elastic
channel. In this section, we will examine only the case $y=0$ and
$L_{y}\rightarrow L$, neglecting such absorptive effects. In this way, the model
is less complicated, nevertheless it explains the suppression (\ref{b=0}). The
general case $y\neq 0$ has been discussed in~\cite{CC09}.

The problem of calculating $S_\el$, by use of the Trotter formula, turns out to
be equivalent to quantize a hamiltonian and to evaluate a matrix element of the
evolution operator $\U(0,\infty)$. In other words, the classical dynamics of the
field $\rho(r^2)$ governed by the lagrangian~(\ref{rhoAction}), is promoted to a
one-dimensional quantum system where $\rho$ plays the role of ``position
variable'' and $r^2\equiv\tau$ represents the ``time'' evolution variable.  The
classical hamiltonian is given by the Legendre transform of $L$ and we quantize
it by imposing canonical commutation relations: introducing the conjugate
momentum $\Pi=\frac{\partial L}{\partial\dot{\rho}}=\frac{1}{2G}(\dot{\rho}-1)$,
we have
\begin{align}
  &H = \Pi\dot{\rho}-L = \frac{1}{4G}\left(
  \dot{\rho}^2-1+\frac{R^2}{\rho}\Theta(\tau-b^2)\right)
  \;, \qquad \tau\equiv r^2
 \label{H} \\
  &[\rho,\Pi] = \ui\hbar \quad\longrightarrow\quad
  \dot{\rho} = -\ui\frac{R^2}{2\alpha}\frac{\partial}{\partial\rho} \;.
 \label{cr}
\end{align}
The Hamiltonian, according to this quantization, is 
\begin{equation}\label{hamquant}
  \frac{\hat{H}}{\hbar}=-\frac{R^2}{4\alpha}\frac{\partial^2}{\partial\rho^2}
   +\alpha\left(\frac{\Theta(\tau-b^2)}{\rho}-\frac{1}{R^2}\right)
  \equiv \frac{H_{0}}{\hbar}+\alpha\frac{\Theta(\tau-b^2)}{\rho}
\end{equation}
In this way, the path integral (\ref{sel}) is equivalent to the matrix element
\begin{equation}\label{S}
  S(b,s)=\bk{\rho=0 |\U(0,+\infty)|\Pi=0}
\end{equation}
where the initial and final states express the boundary conditions for $\rho$
(we recall the relation $\ket{\Pi=0}=\ket{\dot\rho=1}$).
We note that the Hamiltonian is characterized by a Coulomb potential that is
attractive for $\rho<0$ and repulsive in the region $\rho>0$: there is an
infinite Coulomb barrier separating the boundary condition $\rho=0$ from the
(perturbative) region of large $\rho>0$ and by means of this feature we'll be able
to interpret the suppression of the $S$-matrix as a tunnel effect.

The quantum model can be solved for any $b$, but is particularly simple for
$b=0$, where it explains the value $S\sim \esp{-\pi Gs}$ obtained at semiclassical
level.  Indeed, if the impact parameter vanishes, the expression~(\ref{S}) gives
\begin{equation}\label{Sbzero}
  S(b=0,s)=\langle\rho=0 |\U_c(0,+\infty)|\Pi=0\rangle
\end{equation}
where $\U_c$ is the evolution operator related to the time independent Coulomb
hamiltonian $H_c=H_{0}+Gs/\rho$. We calculate the above matrix element by
introducing the Coulomb wave function
\begin{equation}
\psi_c(\rho)=\langle\rho|\U_c(0,\infty)|\Pi=0\rangle 
\end{equation}
so that $S(b=0,s)=\psi_c(0)$. Therefore, the $b=0$ quantum problem is very
similar to the calculation of the Coulomb wave function at the origin for
nuclear processes. It can be shown~\cite{CC09} that, for every $b$,
the $S$-matrix is given by the formula
\begin{equation}
  S(b,s) = \sqrt{\frac{\ui\alpha}{\pi b^2}} \int \dif\rho\;
  \esp{-\ui\alpha(\frac{\rho^2}{b^2}+b^2)}\,\psi_c(\rho)
\end{equation}
which was treated in full details in refs.~\cite{CC08,CC09}.

In order to derive the function $\psi_c(\rho)$, we note that the state
$|\Pi=0\rangle$ is an eigenstate of the free hamiltonian $H_0$ with zero energy,
so that $|\psi_c\rangle$ becomes an eigenstate of the full hamiltonian with null
eigenvalue. Therefore, the wave function can be determined by solving the
stationary Schrodinger equation with zero energy (from now on we use $R=1$ as
unit length)
\begin{equation}\label{schrodinger}
  H_c\,\psi_c = \hbar\left[-\frac{1}{4\alpha}\frac{\dif^2 \psi_c}{\dif\rho^2}
  +\alpha\left(\frac{1}{\rho}-1\right)\psi_c\right] = 0 \;.
\end{equation}
The form of $\psi_c$ is specified, including its boundary conditions, by the
Lippman-Schwinger equation
\begin{equation}\label{lippman}
  \psi_c = \esp{2\ui\alpha\rho}+\alpha G_{0}(0)\,\mathrm{pv}
 \left(\frac{1}{\rho}\right)\psi_c(\rho) 
\end{equation}
with $G(E)=[E-H_{0}-\ui\epsilon]^{-1}$.
\footnote{The $-\ui\epsilon$ prescription is related to the $r^2$-antiordering
  of $\U_c(0,\infty)$.}
For $\rho>0$, where the potential is repulsive, the Coulomb function contains
incident and reflected waves, while it has only a transmitted wave in the region
$\rho<0$ of attractive potential.  This explains why the wave function
$\psi_c(\rho)$ is suppressed in the origin by the tunnel effect through the
Coulomb barrier and the tunneling amplitude gives the order of magnitude of the
$S$-matrix for $b=0$.

Indeed, by solving eq.~(\ref{schrodinger}) with the above boundary conditions,
the wave function $\psi_c$ turns out to be~\cite{CC09}
\begin{equation}\label{funzione}
  \psi_c = \frac{(4\ui\alpha L^2)^{i\alpha}}{\cosh(\pi\alpha)}\,
  z \esp{-\frac{z}{2}}\left[U(1+\ui\alpha,2,z)
  +\frac{\ui\pi\Theta(\ui z)}{\Gamma(\ui\alpha)}F(1+\ui\alpha,2,z)\right]
  \;, \qquad (z\equiv -4\ui\alpha\rho)
\end{equation}
where $L$ is an infrared cutoff that regularizes the Coulomb singularity at large
distances while $U$ and $F$ are the irregular and regular confluent hypergeometric
functions respectively. The value at the origin of the wave function is
\begin{equation}\label{psic0}
  \psi_c(0) = S(0,s) =
  \frac{(4\alpha L^2)^{\ui\alpha}\esp{-\frac{\pi\alpha}{2}}}{\Gamma(1+\ui\alpha)
  \cosh(\pi\alpha)} \sim
  \esp{-\pi Gs} \, \esp{\ui\Re\A_\cl} \left[1+\ord{\frac1{\alpha}}\right]\;.
\end{equation}
In this way we obtain an interpretation for the suppression of the elastic
amplitude for $b<b_c$, as well as quantum corrections to it.

%%%%%%%%%%%%%%%%%%%%%%%%%%%%%%%%%%%%%%%%%%%%%%%%%%%%%%%%%%%%%%%%%%%%%%%%%%%%%%%%
\section{Semiclassical unitarity defect below the critical point\label{s:sud}}
%%%%%%%%%%%%%%%%%%%%%%%%%%%%%%%%%%%%%%%%%%%%%%%%%%%%%%%%%%%%%%%%%%%%%%%%%%%%%%%%

The reduced-action model described in the previous section shows the existence
of a critical impact parameter $b_c\sim R(s)$ such that, for $b<b_c$ the elastic
scattering amplitude is exponentially suppressed with $s$ both at semiclassical
and quantum level, even without inelastic processes ($y=0$). For $b>b_c$ the
small elastic absorption which appears in the quantum model at $y>0$ is expected
to be compensated by emission processes. We do not know about $b<b_c$, but it
might be possible that the elastic suppression due to the tunneling is still
compensated, because of the quantum gravity dynamics.

In any case, the model can be unitary only if we consider the total transition
probability, summed over all emission processes. For instance, at $b=0$,
the suppression factor
\begin{equation}\label{supfac}
  |\bk{0|S|0}|^2 \sim \esp{-2\pi\alpha}
\end{equation}
could be compensated, thus recovering unitarity, only in the sum of all
probabilities
\begin{equation}\label{usum}
   \sum_{n} |\bk{n|S|0}|^2 = \bk{0 | S^\dagger S | 0} \stackrel{?}{=} 1 \;,
\end{equation}
where $\ket{n}$ generically indicates ``the'' state with $n$ gravitons which can
be emitted thanks to the coherent state operator in
eqs.~(\ref{Omega},\ref{d:Sq}).

%===============================================================================
\subsection{Inclusive action and equations of motion\label{s:ia}}
%===============================================================================

In order to study the unitarity properties of our model, we try to evaluate the
vacuum expectation value of $S^\dagger S$. From the definition~(\ref{d:Sq}), it
is clear that the result will be a double path-integral in the fields
$\rho(\tau)$ [from $S$] and $\tilde\rho(\tau)$ [from $S^\dagger$]. The matrix
element between the vacuum states is computed after normal ordering of the
operators $A,A^\dagger$ in the coherent-state operators by means of the
Baker-Campbell-Hausdorff formula
\begin{equation}\label{BCH}
  \bk{0| \esp{-\frac{2\ui\sqrt{\alpha}}{\pi R}\int \dif^2\xt' \;
 [1-\dot{\tilde\rho}(\tau')]\Omega(\xt')}
 \esp{\frac{2\ui\sqrt{\alpha}}{\pi R}\int \dif^2\xt \;
 [1-\dot{\tilde\rho}(\tau)]\Omega(\xt)} |0}
 = \esp{\alpha y \int [\dot{\tilde\rho}(\tau)-\dot\rho(\tau)]^2 \;\dif\tau} \;.
\end{equation}
We obtain the expression
\begin{equation}\label{dpi}
  \bk{0|S^\dagger S|0} = \int\Dif\rho\Dif{\tilde\rho}\;
  \esp{\ui\A_u(\rho,\dot\rho;b)} \;,
\end{equation}
where the usual boundary conditions $\rho(0)=\tilde\rho(0)=0$,
$\dot\rho(\infty)=\dot{\tilde\rho}(\infty)=1$ are understood and we introduced
the {\em inclusive action}
\begin{equation}\label{d:Au}
  \A_u \equiv -\alpha \int_0^\infty \left[
    (1-\dot\rho)^2 - \frac{\Theta(\tau-b^2)}{\rho}
  - (1-\dot{\tilde\rho})^2 + \frac{\Theta(\tau-b^2)}{\tilde\rho}
  -\ui y (\dot\rho - \dot{\tilde\rho})^2 \right] \; \dif\tau \;.
\end{equation}
Here the $y$-dependent term shows the effect due to summing over all
intermediate states.

In the spirit of the semiclassical approximation, a good estimate of the value of
the functional integral for large $\alpha$, i.e., for transplanckian energies,
is obtained by evaluating the inclusive action on the field configuration that
maximizes $\ui\A_u$. The inclusive action is stationary when the fields $\rho$
and $\tilde\rho$ satisfy the Euler-Lagrange equations
\begin{equation}\label{eomU}
  \begin{cases}
    2\ddot\rho - 2\ui y(\ddot\rho - \ddot{\tilde\rho}) &=
    \displaystyle{\frac{\Theta(\tau-b^2)}{\rho^2}} \\[2ex]
    2\ddot{\tilde\rho} + 2\ui y(\ddot{\tilde\rho} - \ddot\rho) &=
    \displaystyle{\frac{\Theta(\tau-b^2)}{\tilde{\rho}^2}}
  \end{cases}
\end{equation}
and the condition of maximum for $\ui\A_u$ must be determined independently for
the various solutions by studying the stability of the action functional for
arbitrary variations of the fields (cfr.~sec.~\ref{s:sirc}).

At the semiclassical level we are considering, it is assumed that the sum over
field configurations represented by the path integral be dominated by just one
solution among those satisfying the Euler-Lagrange equations~(\ref{eomU}).
Therefore, if $\rho$ is such a solution yielding the dominant contribution to
$S$, for symmetry reasons the solution $\tilde\rho$ yielding the dominant
contribution to $S^\dagger$ must be equal to $\rho^*$.  In other words, we argue
that only solutions with $\rho(\tau) = \tilde{\rho}^*(\tau)\equiv \rho_1+\ui\rho_2$ are physically
acceptable within the semiclassical approximation, $\rho_{1,2}$ being real fields.

Once the semiclassical solutions $\rho=\rho_1+\ui\rho_2$ have been determined,
the action is calculated by splitting the integration over $\tau$ in
eq.~(\ref{d:Au}) into two pieces: {\it (i)} the first interval $0\leq\tau\leq b^2$ where
the fields freely evolve and the integrand is constant; {\it (ii)} the second
interval $b^2 < \tau < \infty$ where the evolution is non-trivial but admits a
constant of motion provided by the hamiltonian obtained by Legendre transform of
the lagrangian in eq.~(\ref{d:Au}):
\begin{equation}\label{Hu}
 H_u = 2\dot\rho_1\dot\rho_2 + 2y\dot\rho_2^2 -\frac{\rho_2}{\rho_1^2+\rho_2^2}
 = 0 \;.
\end{equation}
No additional constant of motion exists, so that the system~(\ref{eomU}) is not
solvable analytically. Nevertheless, various relations~\cite{CC09} allow one to
express the inclusive action in terms of few parameters of the solution, namely
the slope $t=t_1+\ui t_2$ of $\rho(\tau)$ in the free region {\it (i)}, and the
asymptotic value $\rinf\equiv\rho_2(\infty)$, according to the
formula~\cite{CC09}
\begin{equation}\label{AuTot}
 \ui\A_u = 4\alpha\left[\rinf - \frac32\frac{t_2}{t_1^2+t_2^2}\right] \;.
\end{equation}
This equation shows that the inclusive action vanishes for real solutions
($\rho_2=0$) in which case the $S$-matrix is unitary. On the contrary,
absorption is present only with a non-vanishing imaginary part $\rho_2$. By
studying the ``inclusive'' equation of motion~(\ref{eomU}), we can determine for
which values of $b$ and $y$ the solutions necessarily develop an imaginary
part, and we can then compute the unitarity defect. However, since no analytic
solution is available, we cannot exhibit an explicit criticality equation, and
we have to resort to numerical or approximate methods.

%===============================================================================
\subsection{Real and complex inclusive solutions\label{s:sirc}}
%===============================================================================

In the previous section we showed that a sufficient condition for unitarity is
the existence of real solutions $\rho=\rho^*=\tilde\rho$, because the inclusive
action vanishes identically, thus implying $\bk{0|S^\dagger S|0}=1$ at
semiclassical level. In this case both equations in~(\ref{eomU}) reduce to the
equation~(\ref{eomrho}) governing the elastic $S$-matrix, with the same boundary
conditions, hence with real solutions only for $b>b_c$. As a consequence, in
this regime the model is unitary. This argument shows that the same critical
point $b_c$ --- found to be $b_c^2=\frac{3\sqrt{3}}{2}R^2$ for the elastic channel at $y=0$
--- also governs the inelastic unitarity of the $S$-matrix in a $y$-independent
way.

\begin{figure}[ht!]
  \centering
  \includegraphics[width=0.35\textwidth]{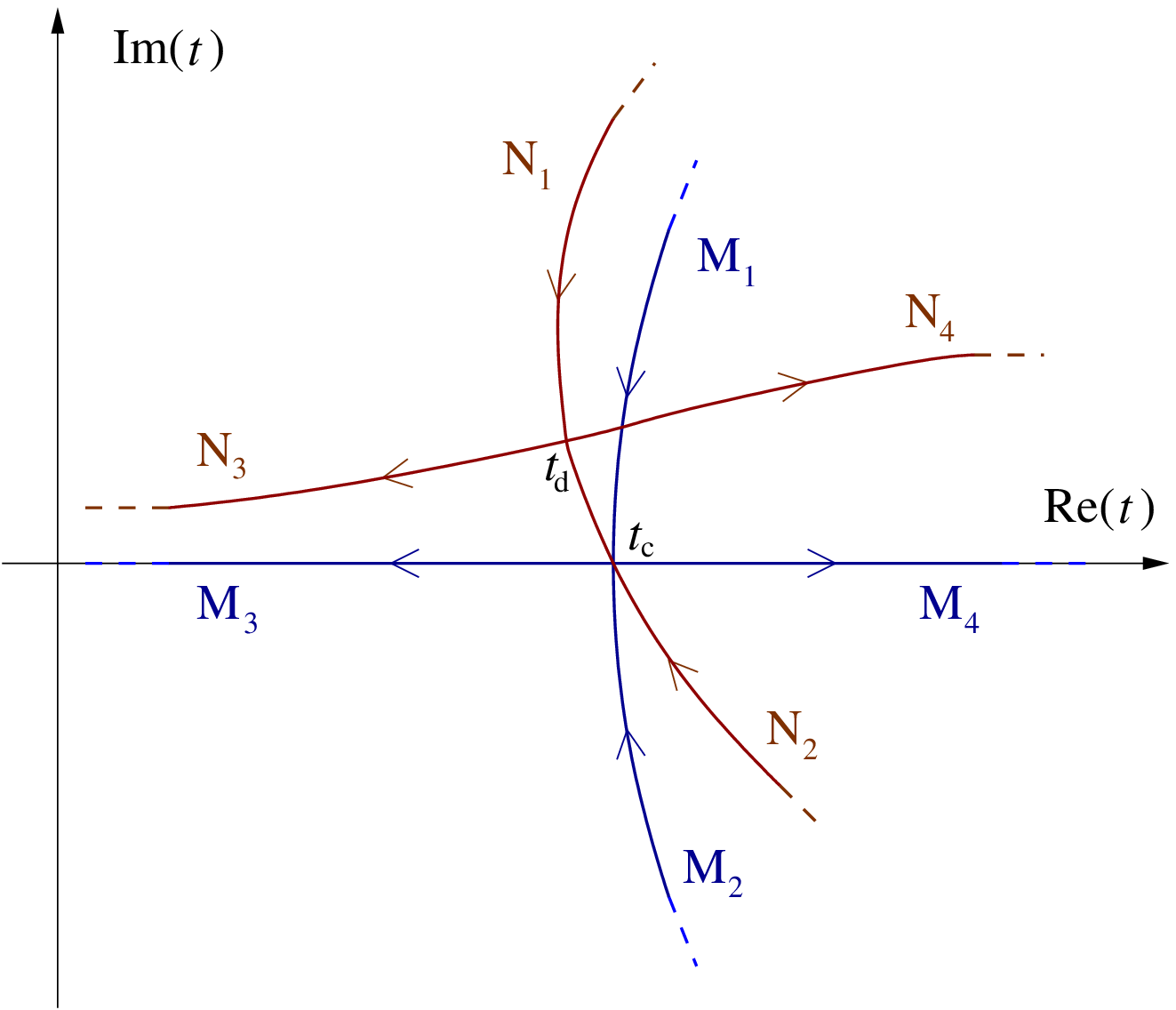}%
  \hspace{0.1\textwidth}%
  \raisebox{0.31\textwidth}{\includegraphics[height=0.5\textwidth,angle=-90]{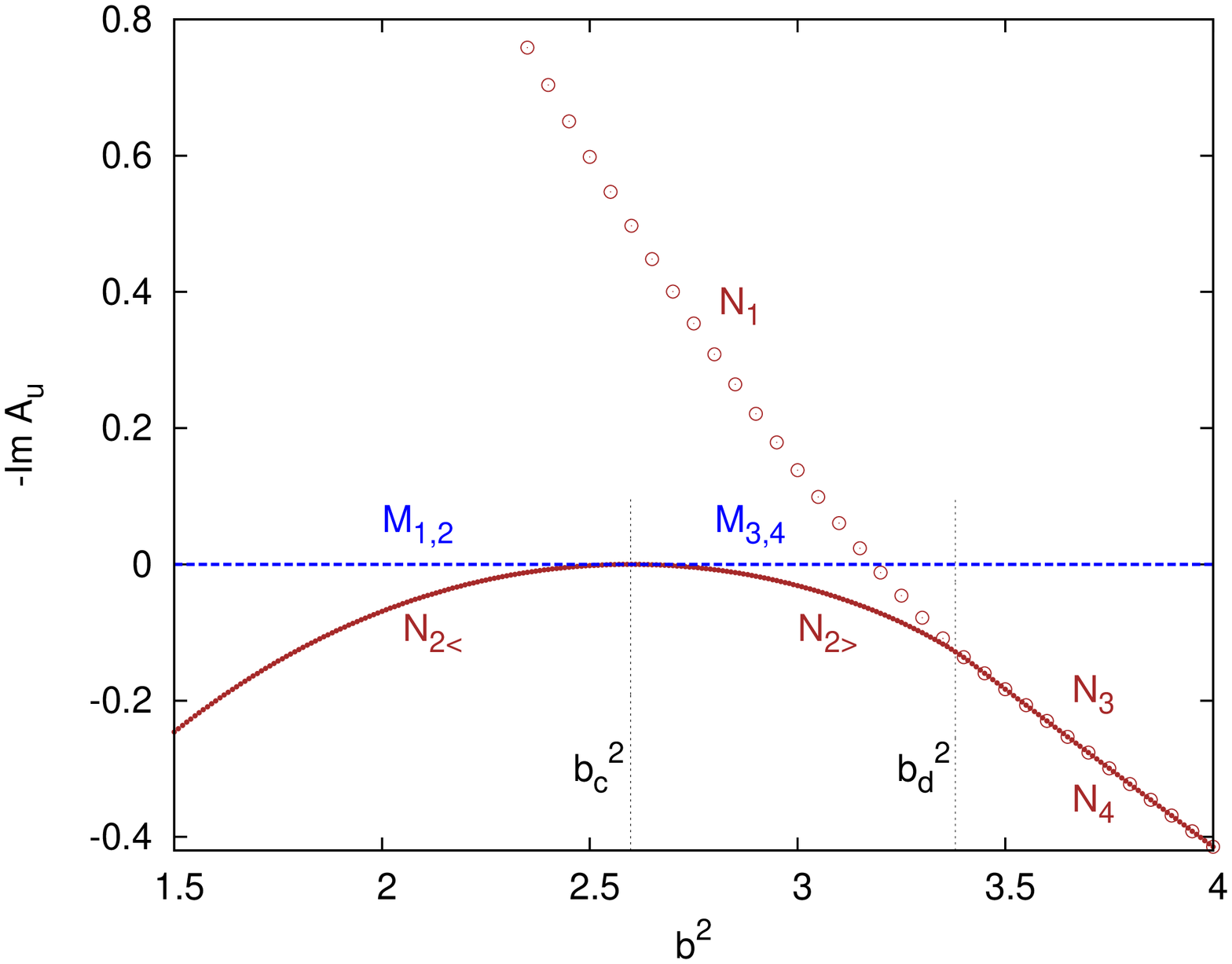}}
  \caption{\it {\em [Left]} Regions of $t$-values spanned by varying $b$ at fixed
    $y=0.5$. {\em [Right]} The inclusive action corresponding to the various solutions.}
  \label{f:crosses}
\end{figure}

The above reasoning also implies that, below the critical impact parameter
($b<b_c$), only genuine complex solutions of~(\ref{eomU}) exist. In order to
characterize such solutions, we have written a numerical program that seeks for
independent pairs $(\rho,\tilde\rho)$ satisfying the inclusive system. We found
that, without imposing the ``physical'' condition $\tilde\rho=\rho^*$, there are
four distinct complex solutions for any given value of $b$ and $y$. By labelling
each solution with the value of $t\equiv\dot\rho(0)$, we can represent the four
solutions with four points in the complex $t$-plane. At fixed $y$ and increasing
$b$ these points move, spanning the curves $M_1\cup M_3$, $M_2\cup M_4$,
$N_1\cup N_3$ and $N_2\cup N_4$ depicted in fig.~\ref{f:crosses}a in the
direction of the arrows. Now, if we consider only the solutions with
$\rho=\tilde\rho^*$, their number varies with $b$ according to the following
scheme:
\begin{itemize}
\item $b<b_c$: there are two complex solutions, one with positive ($N_1$) and one
  with negative ($N_2$) imaginary part;
\item $b_c < b < b_d(y)$: there are two real ($M_3$, $M_4$) and two
  complex ($N_1$, $N_2$) solutions, the latter with positive imaginary part;
here $b_d$ is a $y$-dependent critical value that will be analytically described
in the next section;
\footnote{At $b=b_c$ the three solutions $M_1$, $M_2$ and $N_2$ are pure real
  and coincide. At $b=b_d$ the two solutions $N_1$, and $N_2$ coincide.}
\item $b>b_d(y)$: only the real solutions ($M_3$, $M_4$) survive.
\end{itemize}

As already mentioned, we can accept only field configurations corresponding to
maxima of $\ui\A_u$. Analytically, stability is expressed by requiring the second
variation of $\Im\A_u$ to be positive definite with respect to arbitrary {\em
  real} variations $(\delta\rho,\delta\tilde\rho)$ of the solution.%
\footnote{We consider only real variations of $\rho$ and $\tilde\rho$ for
  compatibility with the definition of the path integral~(\ref{d:Sq}) which is
  defined by integration over real fields.}
By definition, the first-order variation of the inclusive action vanishes on the
solutions: $\delta\A_u = 0$. The imaginary part of the second-order variation yields
\begin{equation}\label{imlag}
  \delta^2 \Im\A_u = \int \left\{ y(\delta\dot\rho-\delta\dot{\tilde\rho})^2
  -\Theta(\tau-b^2)\frac{\rho_2(3\rho_1^2-\rho_2^2)}{(\rho_1^2+\rho_2^2)^3}
  \left[(\delta\rho)^2+(\delta\tilde\rho)^2\right] \right\} \dif\tau \;,
\end{equation}
which is positive definite provided
\begin{equation}\label{posdefcond}
    \rho_2 ( 3\rho_1^2 - \rho_2^2 ) < 0 \;.
\end{equation}
In practice, when $|\rho_2|$ is smaller than $|\rho_1|$, as happens in the
important region $b\sim b_c$, stability occurs for $\rho_2 \leq 0$. Therefore,
the only complex solution satisfying conditions~(\ref{posdefcond}) is found on
$N_2|_{\Im t < 0}$ for $b < b_c$; all other complex solutions are unstable. On the other
hand, any real solution is stable.

To summarize, for $b > b_c$ we have two real and $y$-independent acceptable
solutions which coincide with the exclusive solutions studied in~\cite{CC08},
providing a unitary $S$-matrix. For $b < b_c$ instead, we have only one complex
acceptable solution with negative imaginary part ($\Im(t)<0$) and corresponding
negative action $\ui\A_u < 0$ (cfr.~fig.~\ref{f:crosses}b), thus determining a
unitarity defect for the $S$-matrix.

%===============================================================================
\subsection{Behaviour around the critical point\label{s:bacp}}
%===============================================================================

In order to investigate the onset of absorption below the critical point and to
quantify its magnitude, we study the behaviour of the solutions
of the inclusive equations~(\ref{eomU}) for $b\simeq b_c$. In particular, we
consider the physical ($\rho=\tilde\rho^*$) complex solution $N_2$ which
provides the unitarity defect for $b < b_c$.
The fact that this solution becomes real ($\rho_2(\tau)=0$) for $b=b_c$, suggests
to perform a perturbative analysis in which the imaginary part $\rho_2$ is
considered a small quantity in some neighbourhood of $b\simeq b_c$.

We first set up the relevant equations in a way that is convenient for the
perturbative expansion.  In terms of the real components $\rho_{1,2}$,
eqs.~(\ref{eomU}) read
\begin{subequations}\label{eqs}
  \begin{empheq}[left=\empheqlbrace]{align}
    2\ddot\rho_1 + 4 y \ddot \rho_2 &=
    \Theta(\tau-b^2)\frac{\rho_1^2-\rho_2^2}{(\rho_1^2+\rho_2^2)^2}
    \label{eq1} \\%[3mm]
    2\ddot\rho_2 &=
    \Theta(\tau-b^2)\frac{-2\rho_1 \rho_2}{(\rho_1^2+\rho_2^2)^2}
    \label{eq2} \\%[3mm]
    \rho_1(0) &= 0 \;, \quad \rho_2(0) = 0 \;, \quad \dot\rho_1(\infty) = 1
    \;, \quad \dot\rho_2(\infty) = 0 \;.
    \label{cc}
  \end{empheq}
\end{subequations}
The evolution for $\tau \leq b^2$ is linear in $\tau$:
\begin{equation}\label{trivialEv}
  \rho_{1,2}(\tau) = t_{1,2}\, \tau \;, \qquad\qquad(\tau \leq b^2) \;.
\end{equation}
For $\tau\geq b^2$, thanks to the existence of the integral of
motion~(\ref{Hu}), we can reduce by one the order of the differential
system~(\ref{eqs}), e.g.\ replacing the system~(\ref{eqs}) with
the linear combination (\ref{eq1})$-2y$(\ref{eq2}) and eq.~(\ref{Hu})
itself.

A convenient way of rewriting eq.~(\ref{Hu}) is to consider $\rho_1$ the
independent variable, i.e., $\rho_2 = \rho_2\big(\rho_1(\tau)\big)$. This is
possible since $\rho_1(\tau)$ turns out to be a monotonic increasing function of
$\tau$. By denoting with a prime the derivative with respect to $\rho_1$, we have
\begin{equation}\label{rho2rho1}
 \rho_2' \equiv \frac{\dif\rho_2}{\dif\rho_1} = \frac{\dot\rho_2}{\dot\rho_1} \;.
\end{equation}
Dividing eq.~(\ref{Hu}) by $(\dot\rho_1)^2$ and rearranging the factors, we end
up with an equivalent form of the inclusive equations
\begin{subequations}\label{eomN}
  \begin{empheq}[left=\empheqlbrace]{align}
    2\ddot\rho_1 &=
    \frac{\rho_1^2-\rho_2^2+4y\rho_1\rho_2}{(\rho_1^2+\rho_2^2)^2}
    \label{e1} \\%[3mm]
    2\frac{\rho_2'}{\rho_2} &= \frac1{\dot\rho_1^2(\rho_1^2+\rho_2^2)(1+y\rho_2')}
    \label{e2}% \\[3mm]
  \end{empheq}
\end{subequations}
which is particularly suited for the expansion in $\rho_2$ we are going to
perform.

The crucial observation is that, to each perturbative order in $\rho_2$, there
exists a second integral of motion which allows us to express $\rho_2$ as a
function of $\rho_1$ and of an ``integration constant'',
e.g.~$\rho_2(\infty)\equiv\rinf$. This is easily seen by carrying out explicitly
the calculation.

\paragraph{Lowest order}

By expanding in $\rho_2$ the r.h.s.\ of eqs.~(\ref{eomN}) at lowest
relative order, we find for $\tau > b^2 = b_c^2$
\begin{subequations}\label{eom0}
  \begin{empheq}[left=\empheqlbrace]{align}
    2\ddot\rho_1 &= \frac1{\rho_1^2} %\;, \qquad\qquad\raisebox{-4ex}[0ex]{$(\tau \geq b^2)$}
    \label{eom0a} \\
    2\frac{\rho_2'}{\rho_2} &= \frac1{\dot\rho_1^2\,\rho_1^2} \;.
    \label{eom0b}
  \end{empheq}
\end{subequations}
Eq.~(\ref{eom0a}) is nothing but the exclusive equation~(\ref{eomrho}) which
admits the integral of motion
\begin{equation}\label{Hu0}
  \dot\rho_1^2 + \frac1{\rho_1} = 1 %\qquad\qquad(\tau \geq b_c^2)
\end{equation}
yielding $\dot\rho_1$ as a function of $\rho_1$, i.e.
\begin{equation}\label{drho1zero}
  \dot\rho_1 = \sqrt{1-\frac1{\rho_1}} \;.
\end{equation}

Considering now eq.~(\ref{eom0b}), we can use eq.~(\ref{drho1zero}) to replace
$\dot\rho_1^2$ in the r.h.s., so as to obtain the logarithmic derivative of
$\rho_2$ with respect to $\rho_1$ and, after straightforward integration,
$\rho_2$ itself:
\begin{equation}\label{rho2zero}
 2 \frac{\dif\log\rho_2}{\dif\rho_1} = 2\frac{\rho_2'}{\rho_2} =
 \frac1{\rho_1^2\left(1-\frac1{\rho_1}\right)} \qquad \imp \qquad
 \rho_2^2 = \rinf^2 \left(1-\frac1{\rho_1}\right) \;,
\end{equation}
where we have taken into account the fact that $\rho_1(\infty)=\infty$ and
therefore the integration constant $\rinf$ has to be identified with
$\rho_2(\infty)$.
The matching at $\tau=b^2=b_c^2=3\sqrt{3}/2$ of the above
solutions~(\ref{drho1zero},\ref{rho2zero}) with the free
evolution~(\ref{trivialEv}) provides the lowest order $t$-values
\begin{equation}\label{rinf1}
  t_1 = \frac1{\sqrt{3}} \equiv t_c \;, \qquad
  t_2 = \frac{2}{9} \rinf \;.
\end{equation}

\paragraph{First order}

The first order relative corrections are obtained by expanding eqs.~(\ref{eomN})
to first relative order in $\rho_2$:
\begin{subequations}\label{eom1}
  \begin{empheq}[left=\empheqlbrace]{align}
    2\ddot\rho_1 &= \frac1{\rho_1^2} + 4 y \frac{\rho_2}{\rho_1^3}
    \label{eom1a} \\
    \frac{2\rho'_2}{\rho_2} &= \frac{1-y\rho'_2}{\dot\rho_1^2\,\rho_1^2}
    \label{eom1b}
  \end{empheq}
\end{subequations}
We now substitute the expression~(\ref{rho2zero}) into eq.~(\ref{eom1a}), obtaining
\begin{equation}
  2\ddot\rho_1 = \frac1{\rho_1^2} + 4 y \rinf
  \frac1{\rho_1^3}\sqrt{1-\frac1{\rho_1}}
 \equiv \frac{\dif}{\dif\rho_1} V_1(\rho_1) \;,
\end{equation}
whence the (first-order) conserved quantity
\begin{equation}\label{drho1one}
  \dot\rho_1^2 = 1-[V_1(\rho_1)-V_1(\infty)] = 1-\frac1{\rho_1} + y\rinf
  \frac{8}{15}\left[\sqrt{1-\frac1{\rho_1}}
  \left(2+\frac1{\rho_1}-\frac{3}{\rho_1^2}\right)-2\right]\;,
\end{equation}
where $\dot\rho_1(\infty)=1$ has been imposed.

The first-order correction to $\rho_2$ is now found by substituting
eqs.~(\ref{rho2zero},\ref{drho1one}) into the r.h.s.\ of eq.~(\ref{eom1b}) and,
after integration in $\rho_1$, we obtain
\begin{equation}\label{rho2one}
  \rho_2^2 = \rinf^2
  \left\{1-\frac1{\rho_1}+y\rinf\left[\frac{8}{5}-\frac{8}{3\rho_1}+
  \sqrt{1-\frac1{\rho_1}}\left(\frac{11}{15\rho_1^2}+\frac{28}{15\rho_1}-\frac{8}{5}
  \right) \right] \right\} \;.
\end{equation}

The values of $t_1$ and $t_2$ are again found from the matching at $\tau=b^2$
with the free solution~(\ref{trivialEv}). However, at this level of accuracy,
$b$ is slightly different from $b_c$ and also $t_1$ differs from $t_c$. We
parameterize these differences by introducing the adimensional parameters $\be$
and $\epsilon$ such that%
\footnote{This definition of $\be$ differs at $\ord{\be^2}$ from the analogous
  definition of ref.~\cite{CC08} eq.~(72).}
\begin{equation}\label{d:eta}
  b^2 \equiv \frac{b_c^2}{1-\be} \;, \qquad t_1 \equiv t_c(1+\epsilon) \;.
\end{equation}
Note that $\be > 0$ means $b > b_c$.  Evaluating eqs.~(\ref{drho1one},\ref{Hu},\ref{rho2one}) at
$\tau=b^2$ yields respectively
\begin{align}
  \be &= \Auno y t_2 \;, \qquad \Auno = \frac{4}{5}(9-2\sqrt{3})
  \label{t2one} \\
  \epsilon &= -\frac{\be + \sqrt{3} y t_2}{3} = -\frac1{5}(12-\sqrt{3}) y t_2
 \label{epsone} \\
  \rinf &= \frac{9}{2} t_2 \left[ 1 + \frac{2}{15}(27-\sqrt{3}) y t_2 \right] \;.
 \label{rinf2}
\end{align}

To summarize the results so far, we have found the first-order corrections
to $t_1=t_c(1+\epsilon)$, $t_2$ and $\rinf$, obtaining
\begin{equation}\label{comp1}
  \be \sim \epsilon \sim y t_2 \sim y \rinf \;.
\end{equation}
$\be$ and $\epsilon$ are of the same order, $t_2$ and $\rinf$ are of the
same order but the latter have a $y$ factor relative to the former. In order to
deal also with the $y=0$ case, it is then convenient to write $\be$, $\epsilon$
and $\rinf$ in terms of $t_2$.

\paragraph{Second order}

We can compute the second order quantities by expanding eqs.~(\ref{eomN}):
\begin{subequations}\label{eom2}
  \begin{empheq}[left=\empheqlbrace]{align}
    2\ddot\rho_1 &= \frac1{\rho_1^2} + 4 y \frac{\rho_2}{\rho_1^3} -
  3\frac{\rho_2^2}{\rho_1^4}
    \label{eom2a} \\
    \frac{2\rho'_2}{\rho_2} &= \frac1{\dot\rho_1^2\,\rho_1^2}
    \left(1 - y\rho'_2 + y^2\rho'_2{}^2 - \frac{\rho_2^2}{\rho_1^2}\right) \;.
    \label{eom2b}
  \end{empheq}
\end{subequations}
The explicit calculation is based on the strategy and the results presented at
first order. Here we just write down the relations among the relevant parameters
$\be$, $\rinf$, $\epsilon$ and $t_2$:
\begin{align}
  \be &= A_1 y t_2 - \frac{9}{2} \left(1+ A_2\, y^2 \right) t_2^2 \;, \qquad
  A_2 = \frac{291-112\sqrt{3}}{75}
 \label{beta2} \\
  \epsilon &= -\frac1{5}(12-\sqrt{3}) y t_2 + \frac12
  \left( 1 + \frac{16\sqrt{3}-23}{25} y^2 \right) t_2^2 \;.
  \label{eps2} \\
  \rinf &= \frac{9}{2} t_2 \left[ 1 + \frac{2}{15}(27-\sqrt{3}) y t_2
    + \left ( -3 + \frac{883-96\sqrt{3}}{75} y^2 \right) t_2^2 \right] \;.
  \label{rinf3}
\end{align}

\begin{figure}[!tb]
  \centering
  \includegraphics[width=0.4\textwidth]{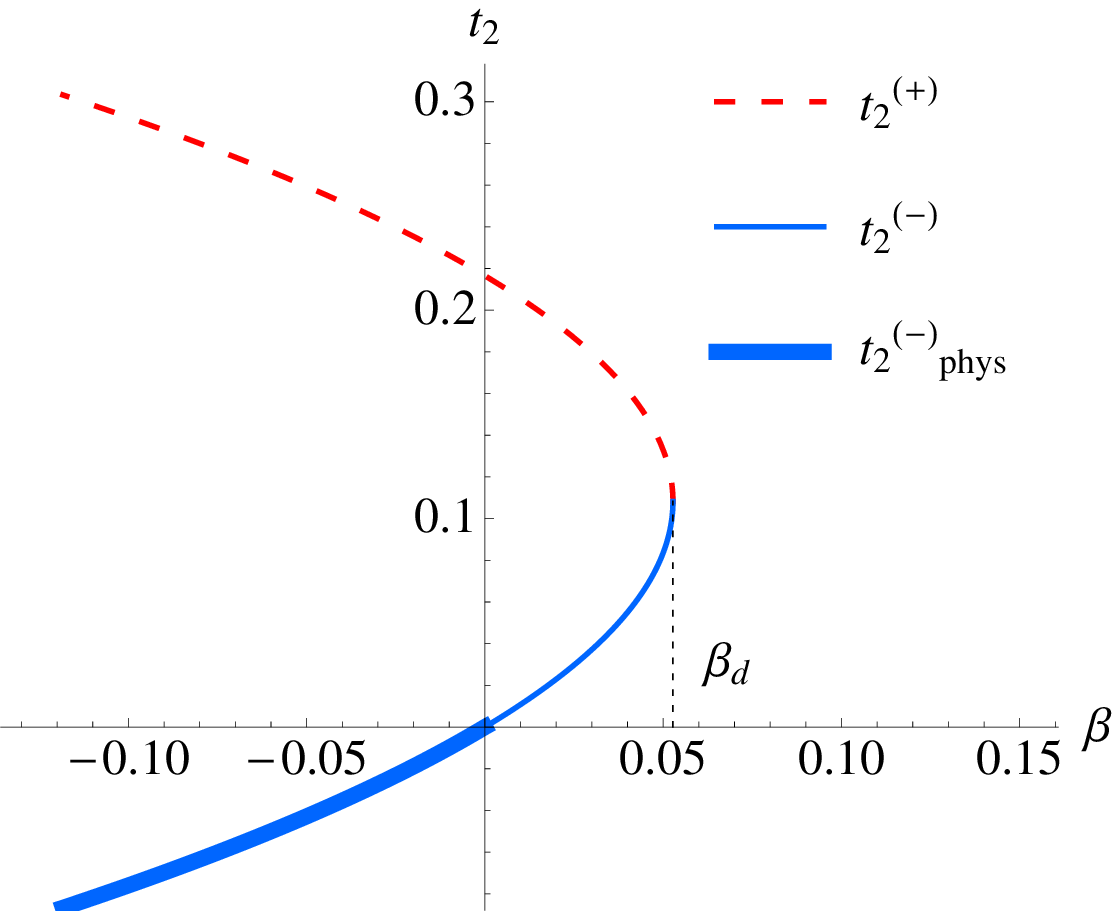}
  \hspace{0.1\textwidth}
  \includegraphics[width=0.4\textwidth]{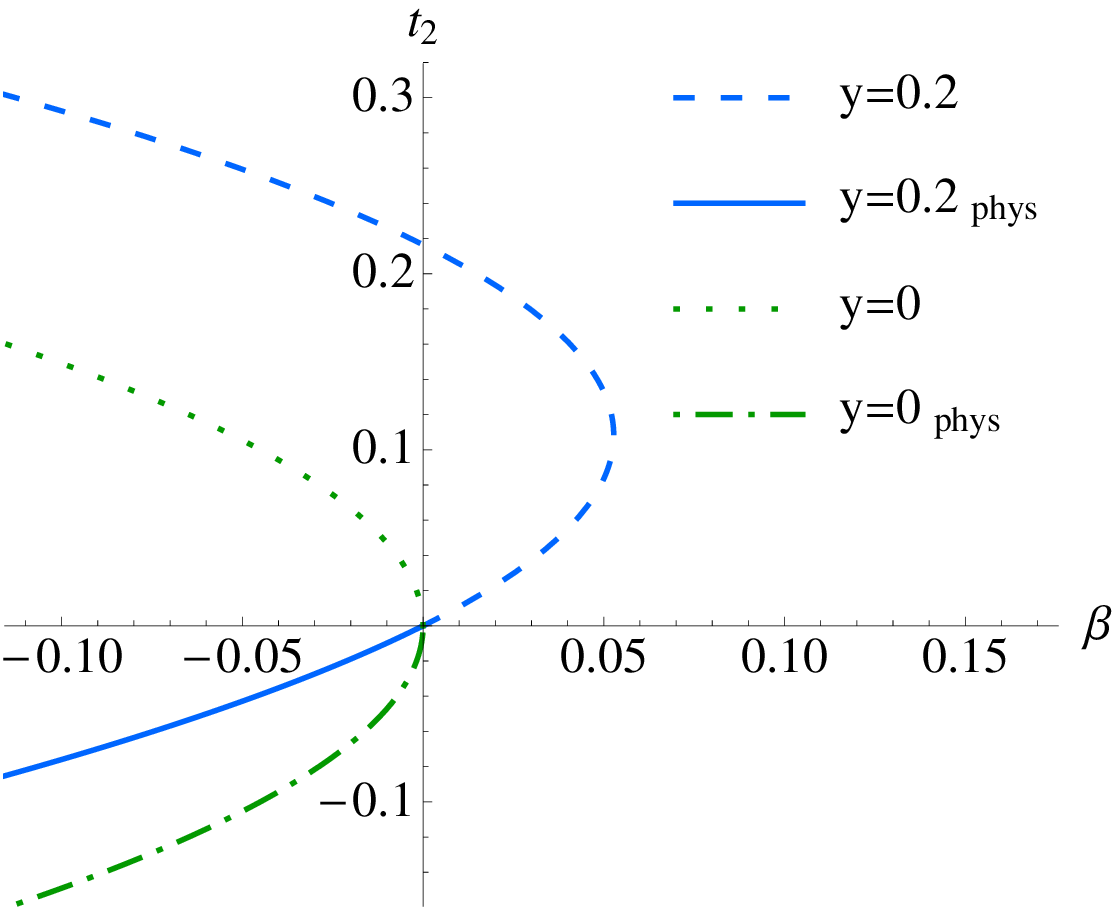}
  \caption{\it {\em [Left]} The two solutions $t_2^{(\pm)}$ (red dashed and blue solid) versus
    $\be$ for $y=0.2$. They coincide at $b=b_d$. Only $t_2^{(-)}$ for $b < b_c$
    is physical (thick line). {\em [Right]} Comparison of the previous solutions for
    $y=0.2$ (blue solid and dashed) with $t_2^{(\pm)}$ in the $y=0$ case (green
    dotted and dash-dotted), the latter showing a square-root behaviour around
    the critical point $\be=0$.}
\label{f:t2vsbeta}
\end{figure}

We note the different behaviour of $t_2$ versus $\be$ at different values of $y$
(fig.~\ref{f:t2vsbeta}b).  Without opened inelastic channels ($y = 0$) the
imaginary component $\rho_2$ of the field grows in a non-analytic way as soon as
$b$ decreases below $b_c$:
\footnote{Actually, the same square-root behaviour occurs in the region
  $y^2 \ll |\be| \lesssim 1$.}
\begin{equation}\label{comp2}
 \rinf \sim t_2 \sim \sqrt{-\be} \sim \sqrt{\epsilon} \;.
\end{equation}
On the other hand, with a finite inelastic emission ($y>0$) the imaginary
component $\rho_2$ grows linearly for small values of $b_c-b$, i.e.,
$|\be| \ll y^2 \lesssim 1$, according to eq.~(\ref{t2one}). In the intermediate region
$y^2\sim|\be|\ll 1$, where both $y$ and $\be$ are small, the quadratic
equation~(\ref{beta2}) yields the two solutions (fig.~\ref{f:t2vsbeta}a)
\begin{equation}\label{t2pm}
 9 t_2^{(\pm)} \simeq A_1 y \pm\sqrt{A_1^2 y^2 -18\be}
\end{equation}
provided its discriminant is greater than zero, and this happens only if the
impact parameter $b$ is smaller than a $y$-dependent critical value $b_d$
determined by
\begin{equation}\label{bd}
  \be \leq \frac{A_1^2 y^2}{18} \equiv \be_d(y) \;, \qquad
  b_d^2(y) \equiv \frac{b_c^2}{1-\be_d} \;.
\end{equation}
The above condition reproduces the structure of the complex solutions
numerically found in sec.~\ref{s:sirc}, namely the existence of the two complex
solutions $N_1$ and $N_2$ for $b_c^2 < b^2 < b_d^2$. For $b > b_c$ both have
positive imaginary part ($t_2^{(\pm)}>0$), while for $b<b_c$ only one of them
($t_2^{(-)}$) becomes negative and is therefore physically acceptable according
to the stability requirement.

Such different behaviours affect the inclusive action. As a consequence,
there are different trends of absorption, according to whether few graviton
emission is allowed ($y\to0$) or a finite contribution of the inelastic channels
is considered.

%===============================================================================
\subsection{Unitarity defect\label{s:ud}}
%===============================================================================

The solutions just obtained with the perturbative method allow us to compute the
unitarity defect $\esp{\ui\A_u}$ --- at least for small enough $y$ and $\be$. In
fact, by recalling eq.~(\ref{AuTot})
\begin{equation}\label{iAu}
 \ui\A_u = 4\alpha\left[\rinf - \frac32\frac{t_2}{t_1^2+t_2^2}\right] \;.
\end{equation}
the imaginary part of the unitarity action is given in terms of $t_1$, $t_2$
and $\rinf$, whose dependence on the impact parameter $b\sim b_c$ has been
analytically obtained in the previous section.

Above the critical point $b\geq b_c$, the physical inclusive solutions are real,
$\rho_2(\tau)$ vanishes identically, and the same is true for $t_2$, $\rinf$ and
$\A_u$, thus implying a unitary $S$-matrix. On the contrary, below the critical
point $b<b_c$, the inclusive action is governed by the complex perturbative
solution $t_2^{(-)} < 0$ which joins continuously with the real solution(s) at
$b=b_c$. In order to study the transition across the critical point, we expand
the action~(\ref{iAu}) in $t_2$ by means of eqs.~(\ref{eps2}) and (\ref{rinf3}),
where we recall that $t_1 = (1+\epsilon)/\sqrt{3}$.

We note the interesting feature that the linear terms in $t_2$ cancel in the
expansion of $\A_u$. Therefore, the suppression of the $S$-matrix starts at
second order in $t_2$:
\begin{equation}\label{act3}
  \frac{\ui\A_u}{\alpha} = -\frac{12}{5}(9-2\sqrt{3}) y t_2^2 +
  18 \left[ 1 - \frac{509-168\sqrt{3}}{75} y^2 \right] t_2^3 + \cdots
\end{equation}
Due to the interplay between $y t_2^2$ and $t_2^3$, $\A_u$ is characterized by
various regimes.

At $y=0$ the action~(\ref{act3}) is cubic in $t_2 \sim -\sqrt{-\be}$:
\begin{equation}\label{Auy0}
  \frac{\ui\A_u}{\alpha} \simeq 18 t_2^3 \simeq -\frac{4\sqrt{2}}{3}
  (-\be)^{3/2} 
  = -\frac{2}{\alpha} \Im \A_\el \;,
\end{equation}
and reproduces the fractional exponent $3/2$ of the elastic action~(\ref{A}) as
shown in ref.~\cite{ACV07}. Note that $\ui\A_u$ is negative due to the choice of
the stable solution $t_2^{(-)} < 0 $. As a consequence, the $S$-matrix is
suppressed for $b<b_c$. The same happens at small
$y \ll |t_2| \sim \sqrt{|\be|}$.

At intermediate $y\sim |t_2| \ll 1$, the two terms in eq.~(\ref{act3}) are of the
same order and the $S$-matrix is always suppressed for $b < b_c$ since
$t_2^{(-)} < 0$.

At finite $y \gg \sqrt{|\be|}$, the term proportional to $y t_2^2$ dominates the
inclusive action, which shows a negative-definite quadratic behaviour in
$b-b_c$, with a vanishing maximum at the critical point:
\begin{equation}\label{act2}
  \frac{i\A_u}{\alpha} \simeq -\frac{12}{5}(9-2\sqrt{3}) y t_2^2
  \simeq -\frac{3}{A_1} \frac{\be^2}{y} \;.
\end{equation}
This is in agreement with the numerical result represented by the solid line in
fig.~\ref{f:crosses}b.

The above analysis for the unitarity action can be summarized by distinguishing
three regimes, at least for $y \lesssim 1$:
\begin{itemize}
\item For $b<b_c$ and $(b-b_c)^2 \gg y$, the unitarity action $\ui\A_u$ is
  negative and decreases in size with a power-like behaviour
  $(b_c-b)^{3/2}$ as $b\to b_c$. This includes the case $y=0$.
\item For $b<b_c$ and $(b-b_c)^2 \ll y$, the unitarity action $\ui\A_u$ is
  negative and vanishes quadratically at $b=b_c$.
\item For $b\geq b_c$ the unitarity action vanishes identically.
\end{itemize}
It is important to realize that the proper choice of the physical solution
avoids a unitarity excess $\bk{0|S^\dagger S|0} > 1$. Such an unphysical
behaviour would occur if we don't reject the $t_2^{(+)}$ solution, since the
corresponding inclusive action $\ui\A_u(t_2^{(+)})$ would be positive for
$b\lesssim b_c$, as shown by the circles in fig.~\ref{f:crosses}b.

\begin{figure}[hb!]
  \centering
  \includegraphics[angle=270,width=0.6\textwidth]{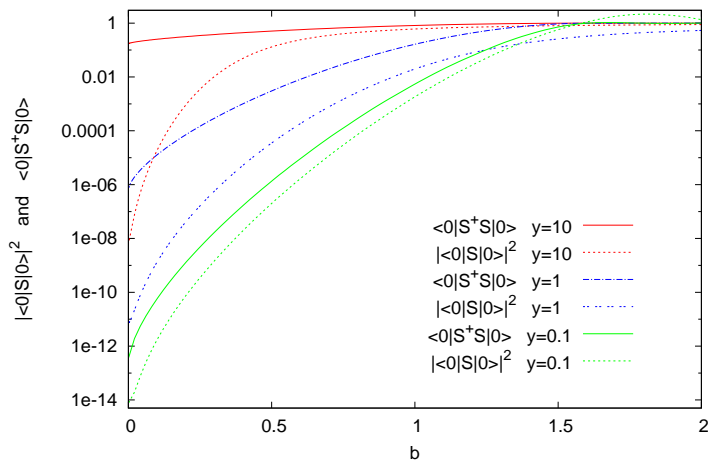}
  \caption{\it Numerical evaluation of the unitarity deficit (solid lines) at various values
    of $y$, and comparison with the quantum v.e.v.~squared of the $S$-matrix
    (dashed lines) illustrating the increasing contribution of the
    inelastic channels at larger $y$.}
  \label{f:osso}
\end{figure}

The numerical evaluation of the ensuing $S$-matrix suppression for various
values of $y$ is presented in fig.~\ref{f:osso}, where we plot the semiclassical
estimate of the v.e.v.~$\bk{0|S^\dagger S|0}=\esp{\ui\A_u}$ and compare it with
the $n=0$ vacuum contribution $|\bk{n=0|S|0}|^2$ of the unitarity
sum~(\ref{usum}). The results show that, if $y$ is small, there is a considerable
unitarity defect in the $S$-matrix, the inelastic channels providing only a small
correction to the elastic suppression. At larger values of $y$ there is an
important recovery of unitarity due to the contribution of the inelastic
channels. However, it must be kept in mind that $y$ --- related to the maximum
rapidity of the emitted gravitons --- cannot be arbitrarily large, because
energy conservation limits the energy of the emitted particles to be smaller
than the available energy $\sqrt{s}$. Actually, energy conservation prevents $y$
to assume large values, which at most can be of order $y\sim\ord{1}$.

In conclusion, a unitary defect is always present in the semiclassical estimate
of $\bk{0|S^\dagger S|0}$ when the impact parameter $b$ is smaller than the
critical one $b_c$. In this region the contribution of inelastic processes only
partially compensates the suppression of the elastic channel, at least at
semiclassical level.

%%%%%%%%%%%%%%%%%%%%%%%%%%%%%%%%%%%%%%%%%%%%%%%%%%%%%%%%%%%%%%%%%%%%%%%%%%%%%%%%
\section{Search for quantum transitions\label{s:sqt}}
%%%%%%%%%%%%%%%%%%%%%%%%%%%%%%%%%%%%%%%%%%%%%%%%%%%%%%%%%%%%%%%%%%%%%%%%%%%%%%%%

The semiclassical estimate of unitarity of the $S$-matrix performed in the
previous section gave an exponential suppression of
$\bk{0|S^\dagger S|0}\sim\esp{-Gs}$ for $b<b_c$. In this section we
investigate whether quantum effects provide larger contributions eventually
restoring unitarity. The quantity that we are going to study at quantum level is
the eigenvalue of the $S$-matrix operator~(\ref{d:Sq}). This is possible because
the specific form of the $S$-matrix --- a coherent-state operator acting on the
Fock space of gravitons.  --- allows us to determine its eigenvectors and
eigenvalues~\cite{CC09}, as we briefly recall.

In order to construct such eigenvectors, it is convenient to introduce
(normalized) graviton-coherent-states which are parameterized by arbitrary
(complex) profile functions $\eta(\tau)$ as follows:
\begin{equation}
 \ket{\eta(\tau)} \equiv
 \esp{-\frac12(\eta^*,\eta)} \esp{(\eta^*,a^\dagger)} \ket{0} \;,
\end{equation}
with the short-hand notation
$(\eta,\zeta)\equiv\int_0^\infty\eta(\tau)\zeta(\tau)\,\dif\tau$ in terms of the
azimuthally-averaged annihilation operators of eq.~(\ref{Omega})
\begin{equation}\label{azimA}
 a(\tau=\xt^2) \equiv \int_0^{2\pi} \frac{\dif\phi_{\xt}}{2\sqrt{\pi}} \;
 \frac{A(\xt)}{\sqrt{Y}} \qquad\Rightarrow\qquad
 [a(\tau),a^\dagger(\tau')] = \delta(\tau-\tau') \;.
\end{equation}
Since the action of the $S$-matrix operator on those states is a superposition
of coherent states with shifted parameter
\begin{equation}\label{Scs}
  S\ket{\eta} = \int\Dif\rho\; \esp{-\ui\int L(\rho)\;\dif\tau}
 \, \ket{\eta+\ui\delta_\rho} \;, \qquad
 \left(\delta_\rho \equiv \sqrt{2\alpha y}(1-\dot\rho)\right) \;,
\end{equation}
the functional Fourier transform of coherent states with imaginary parameter
\begin{equation}\label{funcFT}
 \Ket{\omega(\tau)} \equiv \esp{\frac14(\omega,\omega)}
 \int\Dif{\zeta(\tau)} \; \esp{-\ui(\zeta,\omega)} \, \ket{\ui\zeta(\tau)}
\end{equation}
constitutes a complete set of $S$-matrix eigenstates. With the pre-factor
$\esp{\frac14(\omega,\omega)}$ such states are normalized according to the delta
functional
\begin{equation}\label{omegaNorm}
 \bk{\{\omega'(\tau)\}|\{\omega(\tau)\}} = \delta(\{\omega-\omega'\}) \;.
\end{equation}
By acting on the states~(\ref{funcFT}) with the $S$-matrix~(\ref{d:Sq}), it is
straightforward to verify that the former are eigenstates of the latter. The
corresponding eigenvalues, that we denote with $\esp{\ui\A[\omega]}$, obviously
depend on the real functional parameter $\omega(\tau)$, and are given by the
path-integral
\begin{equation}\label{eig}
  \text{eigenv}_\omega(S) \equiv \esp{\ui\A[\omega]}
  = \int_{\rho(0)=0}^{\dot\rho(\infty)=1} \Dif{\rho} \;
  \esp{-\ui\int L(\rho)+\ui(\delta_\rho,\omega)} \;.
\end{equation}

Endowed with eigenstates and eigenvalues of the $S$-matrix, we can perform a
quantitative study of its unitarity properties at quantum level. For instance,
we can reconsider the v.e.v.\ of $S^\dagger S$ by inserting the completeness of
$\omega$ states, obtaining
\begin{equation}\label{oSSo}
  \bk{0|S^\dagger S|0} =
  \int\Dif{\omega} \; \bk{0|S^\dagger|\{\omega\}}\bk{\{\omega\}|S|0}
  = \int\Dif{\omega} \; \esp{-\frac12(\omega,\omega)} \esp{-2\Im\A[\omega]} \;.
\end{equation}
In the r.h.s.\ of the previous expression, we note two elements that determine
the order of magnitude of the v.e.v.: the ``density'' of the intermediate state
$(\omega,\omega)$ and the eigenvalue (actually, its modulus squared).  If we
find $\Ket{\omega}$ states such that $(\omega,\omega)=\ord{1}$ and
$\Im\A[\omega]$ is much smaller than the tunneling exponent, we have a case for
an important quantum effect, reducing the unitarity defect.

At semiclassical level, the eigenvalue~(\ref{eig}) is found from the
contribution of $\rho$ which makes the action
\begin{equation}\label{omegaAct}
 \A[\omega] \equiv -\ui\int\left[L(\rho,\dot\rho,\tau)
 - \sqrt{2\alpha y} (1-\dot\rho)\omega \right] \; \dif\tau
\end{equation}
stationary. The ensuing Euler-Lagrange equation is similar to that of the
elastic amplitude~(\ref{eomrho}), but with an additional term representing an
``external force'' proportional to $\dif\omega/\dif\tau$ which depends on the
eigenstate:
\begin{equation}\label{omegaEq}
  \ddot\rho -\frac{\Theta(\tau-b^2)}{2\rho^2}
  = -\sqrt{\frac{2y}{\alpha}} \dot\omega(\tau) \;.
\end{equation}
It is this external force that might help to cross the repulsive Coulomb barrier
thus avoiding the exponential suppression for proper values of $\omega(\tau)$.

Here, however, we want to estimate the eigenvalue~(\ref{eig}) at quantum level.
In order to do that, we note that $\esp{\ui\A[\omega]}$ can be rewritten as a
matrix element --- in the Hilbert-space quantizing the 1D system~(\ref{omegaEq})
--- of a proper evolution operator $\U_\omega(0,\infty)$, in complete analogy to
the quantum analysis of the elastic amplitude of sec.~\ref{s:qsm}:
\begin{equation}\label{0U1}
  \esp{\ui\A[\omega]} = \bk{\rho=0|\U_\omega(0,\infty)|\dot\rho=1} \;.
\end{equation}
The bra and ket states embody the boundary conditions of the functional
integration, while the M\"oller operator $\U_\omega$ governs the
$\omega(\tau)$-dependent dynamics, and is determined in the following way.
Starting from the effective lagrangian defined by the integrand in
eq.~(\ref{omegaAct}), by Legendre transform we derive the conjugate momentum and
the hamiltonian
\begin{align}
  \Pi &\equiv \frac{\partial L_\omega}{\partial\dot\rho} =
  2\alpha(\dot\rho-1)+\sqrt{2\alpha y} \, \omega(\tau)
  \label{omegaCC} \\
  H_\omega & =
  \alpha\left[\left(\frac{\Pi+2\alpha-\sqrt{2\alpha y}\,\omega}{2\alpha}\right)^2
    +\frac{\Theta(\tau-b^2)}{\rho}\right] + \sqrt{2\alpha y}\,\omega \;.
\end{align}
We quantize this system by imposing canonical commutation relations
$[\rho,\Pi] = \ui$, e.g., by identifying the operator
$\Pi+2\alpha = -\ui\partial/\partial\rho$ in the $\rho$-coordinate
representation. Finally, the evolution operator $\U_\omega$ obeys the
differential equation
\begin{equation}\label{Ueq}
  \ui\frac{\partial}{\partial\tau} \U_\omega(\tau,\infty) = H_\omega \U_\omega \;.
\end{equation}

By defining the state
\begin{equation}\label{psiOmega}
  \ket{\psi_\omega(\tau)} \equiv \U_\omega(\tau,\infty)\ket{\dot\rho=1}
  \qquad\text{obeying}\qquad
  \ui\frac{\partial}{\partial\tau}\ket{\psi_\omega}
  = H_\omega \ket{\psi_\omega} \;,
\end{equation}
we can express the $S$-matrix eigenvalue~(\ref{0U1}) as the amplitude of the
wave function at $\rho=0$ and $\tau=0$:
\begin{equation}\label{psi00}
  \psi_\omega(\rho,\tau) \equiv \bk{\rho|\psi_\omega(\tau)} \qquad
 \imp \qquad \esp{\ui\A[\omega]} = \psi_\omega(0,0) \;.
\end{equation}
The above wave function can be determined by solving the Schr\"odinger
equation~(\ref{psiOmega}b). Before doing that, we eliminate the
$\omega(\tau)$-dependent shift in momentum and energy by the
similarity transformation
\begin{equation}\label{simTr}
  \psi_\omega(\rho,\tau) \to \esp{\ui\sqrt{2\alpha y}\,\omega(\tau)\rho} \;
  \esp{\ui\sqrt{2\alpha y}\,\int_\tau^\infty\omega(\tau')\;\dif\tau'} \;
  \Psi_\omega(\rho,\tau)
\end{equation}
so that the Schr\"odinger equation for $\Psi$ is
\begin{equation}\label{schr2}
 \ui\frac{\partial}{\partial\tau}\Psi(\rho,\tau) = \left[\alpha
 \left(-\frac1{4\alpha^2}\frac{\partial^2}{\partial\rho^2} -1
   + \frac{\Theta(\tau-b^2)}{\rho} \right) + \sqrt{2\alpha
   y}\,\rho\,\dot\omega(\tau)
  \right] \Psi(\rho,\tau)
\end{equation}
The shift in momentum has produced a linear potential term due to a
$\tau$-dependent external force, as already noted before eq.~(\ref{omegaEq}).
Since $\psi$ and $\Psi$ differ only by an unimportant phase factor, we can
replace the former with the latter in eq.~(\ref{psi00}) and use the hamiltonian
in eq.~(\ref{schr2}) to compute the time-evolution operator $\U_\omega$.

Let us now compute $\Psi(0,0)$ in the case $b=0$: in this way the
$\tau$-dependence only comes from the external force $\dot\omega$. We adopt the
perturbative approach by splitting $H_\omega$ in an unperturbed term
$H_{\omega=0}=H_c$ which coincides with the ``Coulomb''
hamiltonian~(\ref{schrodinger}) governing the elastic amplitude, and in a
perturbation given by the time-dependent potential $\sqrt{2\alpha
  y}\dot\omega\rho$.  For the sake of simplicity, we assume that the external
force is active only within a finite time range $0<\tau<T$ so that for $\tau >
T$ the evolution is just of Coulomb type.  In this way the wave function can be
determined by expanding the evolution operator $\U_\omega$ in Dyson's series,
yielding at first order
\begin{equation}\label{psiPert}
  \Psi_\omega(\rho,0) = \psi_c(\rho) - \ui \sqrt{2\alpha y}
  \int_0^T \dif\tau \; \dot\omega(\tau) \,
  \bk{\rho|\esp{\ui H_c \tau} \rho \,\esp{-\ui H_c \tau} \;|\psi_c} + \cdots \;,
\end{equation}
We obtain a more explicit expression by inserting a complete set of unperturbed
eigenstates $\ket{\phi^{(n)}}$ of energy $E_n$ before the operator $\rho$, by exploiting
eq.~(\ref{schrodinger}), i.e., $H_c\ket{\psi_c}=0$, and then projecting on the
position eigenstate $\bra{\rho=0}$:
\begin{equation}\label{pertRes}
  \esp{\ui\A[\omega]}|_{b=0} \simeq \psi_c(0) - \ui \sqrt{2\alpha y} \;
  \sum_n \phi^{(n)}(0) \; \bk{\phi^{(n)}| \rho | \psi_c}
  \int_0^T \esp{\ui E_n \tau} \dot\omega(\tau) \; \dif\tau \;.
\end{equation}
We see that the effect of the interaction potential $\sim\dot\omega\rho$ is to
cause transitions from the initial state $\psi_c$ towards other eigenstates of
$H_c$.  Eq.~(\ref{pertRes}) can then be interpreted by saying that quantum
effects provide a recovery of unitarity if the interaction potential induces
transitions towards states with a wave function at the origin much larger than
the tunneling amplitude: $\phi_n(0) \gg \psi_c(0)\sim\esp{-\pi\alpha}$. We
expect that such states could belong to two groups:
\begin{itemize}
\item bound states whose wave function is significantly different from zero in
  the region $\rho < 0$ and is finite at $\rho=0$;
\item high-energy continuum states for which the suppression from the Coulomb
  barrier is smaller.
\end{itemize}
More precisely, the entity of the contribution of those states to the eigenvalue
$\esp{\ui\A[\omega]}$ in eq.~(\ref{pertRes}) is given by three factors: the wave
function at the origin $\phi_n(0)$, the matrix element $\bk{\phi_n\rho|\psi_c}$
and the time integral. The first two will be studied in detail in the next
sections. As for the time integral, we need to specify the function
$\omega(\tau)$. By analogy with the quantum mechanical phenomenon of induced transitions, we
consider as a good candidate of external force an oscillatory function which
can induce big transition amplitude at resonance. More precisely we choose
\begin{equation}\label{omegaRes}
  \omega(\tau) = \frac1{\sqrt{T}} \sin(K\tau) \Theta(T-\tau) \;,
\end{equation}
where the normalization has been chosen to keep $(\omega,\omega)=\ord{1}$ in
order to avoid the exponential suppression in eq.~(\ref{oSSo}). Whith this
choice, the time-factor reads
\begin{equation}\label{tint}
 {\cal T}_n \equiv \int_0^T \esp{\ui E_n \tau} \dot\omega(\tau) \; \dif\tau =
 \frac{K}{2\sqrt{T}}\left[\frac{\esp{\ui(E_n-K)T}-1}{\ui(E_n-K)} +
   (K\to-K)\right] \;.
\end{equation}
We are interested in situations where the perturbative corrections to the
eigenvalue~(\ref{pertRes}) are larger than the lowest order term. In this case,
the square modulus of the eigenvalue can be written as
\begin{equation}\label{eigSq}
  \left|\esp{\ui\A[\omega]} \right|^2 \simeq
  \sum_{n,n'} c_n c_{n'}^* {\cal T}_n {\cal T}_{n'}^*
  \simeq K^2 \sum_{n,n'} c_n c_{n'}^* \esp{\ui(E_n - E_{n'})T}
  \frac{\sin[(E_n-K) T]\sin[(E_{n'}-K)T]}{(E_n-K)(E_{n'}-K) T} \;,
\end{equation}
where the $c_n$'s contain the $\rho$ matrix elements and the wave function at
the origin. For long interaction time $T$, the preceding expression is peaked
for $E_n=K$ and provides a delta function of energy conservation
$K^2 \delta(E_n-K)$ which suppresses the interference terms with respect to the
squares of the individual amplitues. The latter provide a contribution of order
$\ord{1}$ around the states compatible with conservation. Therefore, the time
factor is $\ord{1}$ if $(\omega,\omega)=\ord{1}$.

%===============================================================================
\subsection{Transitions to bound states\label{s:bs}}
%===============================================================================

According to eq.~(\ref{pertRes}), the entity of quantum effects to unitarity
depends crucially on the matrix elements $\bk{\phi_n|\rho|\psi_c}$ and on the
wave function at the origin $\phi_n(0)$ of the state reached by the induced
transition. In this section we determine both factors for the case of
transitions to bound states.

In fact, the Coulomb potential $\sim 1/\rho$ in the unperturbed hamiltonian
$H_c$ is attractive for $\rho<0$ giving rise to a discrete spectrum associated
to bound states which could correspond to collapsed states characterized by a
strong gravitational field $h\propto(1-\dot\rho)$. The discrete spectrum of
$H_c$ is given by the ``energy'' eigenvalues
\begin{equation}\label{bse}
  E_n = -\alpha-\frac{\alpha^3}{(n+\frac12)^2} \qquad (n\in\N)
\end{equation}
and the (normalized) eigenfunctions are expressed in terms of the irregular
Whittaker function~\cite{AS} $W$ (closely related to the confluent
hypergeometric function $U$)
\begin{align}
  \phi_n(\rho) &= N_n\left[
    \Gamma\left(\threehalf+n\right) W_{-(n+\half),\half}(x)\;\Theta(x)
  + \Gamma\left(\threehalf-n\right) W_{n+\half,\half}(-x)\;\Theta(-x)
 \right]
 \label{bsf} \\
  N_n &= \phi_n(0) = \frac{c\,\alpha}{(n+\frac12)^{3/2}} \;, \qquad
  x \equiv \frac{4\alpha^2\rho}{n+\half}
 \label{bsn}
\end{align}
where $c \simeq 0.45$. These wave functions are significantly different from
zero for $\rho<0$, and in particular where the energy is larger than the
potential (see fig.~\ref{f:psicpsin}). At variance with
$\psi_c(0)\sim\esp{-\pi\alpha}$, their value at the origin $\phi_n(0)$ is not
suppressed with $\alpha$, but only by a power of $n$, as can be read from
eq.~(\ref{bsn}). For this reason, transitions to bound states might be important
for the estimate of the eigenvalue~(\ref{pertRes}).  The crucial factor is then
the matrix element
\begin{equation}\label{bsme}
 \bk{\phi_n| \rho | \psi_c} = \int_{-\infty}^{+\infty}
 \phi_n^*(\rho) \rho \, \psi_c(\rho) \;\dif\rho 
\end{equation}
which can be estimated by using the WKB approximations for $\phi_n$ and
$\psi_c$.  We use the notation $\nplus\equiv n+\half$ and divide the integration
domain into 3 regions:

\begin{figure}[ht!]
  \centering
  \includegraphics[width=0.6\linewidth]{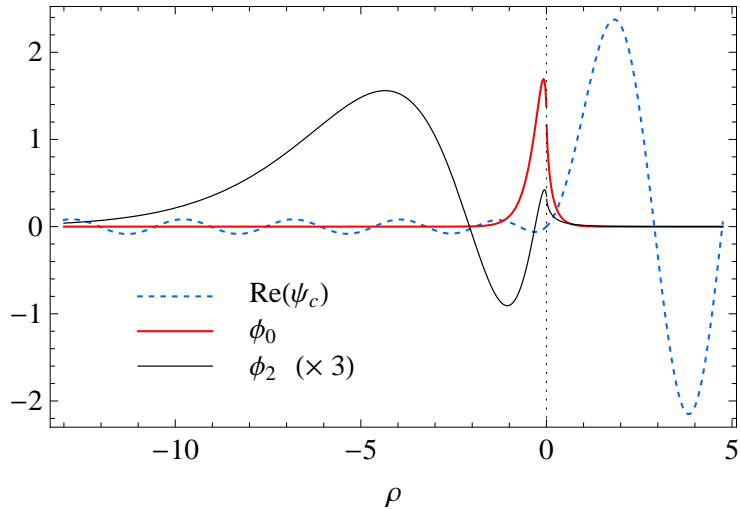}
  \caption{\it Plot of $\Re \psi_c$ (dashed) and of $\phi_n$ (solid) for $n=0$
    (thick) and $n=2$ (thin), the last function being multiplied by a factor of
    3. Here $\alpha = 1$ is small in order not to have a huge suppression of
    the wave functions in the respective classically forbidden regions.}
  \label{f:psicpsin}
\end{figure}

\paragraph{$\bs{(\rho < 0)}$} For negative $\rho$, $\psi_c$ oscillates with
amplitude $\simeq\esp{-\pi\alpha}$ because of the tunneling suppression. The
generic bound state eigenfunction oscillates in the interval
$-(\nplus/\alpha)^2 < \rho < 0$, reaching its maximum value in the leftmost
half-period, with amplitude $\sim \alpha \nplus^{-7/3}$, and then decreases
exponentially towards zero for $\rho\to-\infty$. Therefore, the convergent
oscillatory integrand is uniformly bounded by
$\alpha^{-1} \nplus^{-1/3} \esp{-\pi\alpha}$ and the ensuing contribution to the
matrix element is exponentially suppressed in $\alpha$.

\paragraph{$\bs{(\rho>1)}$} In this region, $\psi_c$ is an oscillating function
with maximum amplitude $\sim\alpha^{1/6}$. On the other hand,
$|\phi_n| < N_n \Gamma(1+\nplus)\nplus^{-\nplus} \esp{-2\alpha^2\rho/\nplus}$ is
strongly suppressed. It turns out that the product $\phi_n \psi_c$ is
oscillatory and exponentially suppressed in $\alpha^2$, and so is the
contribution of this region to the matrix element~(\ref{bsme}).

\paragraph{$\bs{(0<\rho<1)}$} In the intermediate region $|\psi_c|$ increases
from the value $|\psi_c(0)| \simeq \esp{-\pi\alpha}$ to values of order one,
like $\esp{-\pi\alpha} \esp{4\alpha\sqrt\rho}$, while $\phi_n$ starts from
$\phi_n(0)\sim\alpha/\nplus^{3/2}$ and goes to zero as
{\it (i)} $N_n \esp{-4\alpha\sqrt\rho}$ if $\rho \ll (\nplus/\alpha)^2$;
{\it (ii)} 
$N_n\Gamma(1+\nplus)(x+\nplus)^{-\nplus}\esp{-2\alpha^2\rho/\nplus}$
if $\rho \gtrsim (\nplus/\alpha)^2$. It turns out that the product of wave
functions $\phi_n\psi_c$ is also in this case exponentially suppressed. In fact,
in case {\it(i)}, the two esponential with $\sqrt{\rho}$ cancel out and the
product is of order $\esp{-\pi\alpha}$; in case {\it (ii)} we have
\begin{equation}\label{psicpsin}
  |\phi_n \psi_c|\lesssim \esp{-\pi\alpha} N_n \sqrt{2\pi\nplus}\esp{-\nplus}
  \esp{f(x,\nplus)} \;, \quad f(x,\nplus) \equiv -\frac{x}{2}+2\sqrt{\nplus x}
  -\nplus\ln\left(1+\frac{x}{\nplus}\right) \;, \quad \nplus\equiv n+\frac12 \;.
\end{equation}
The exponent $f(x,\nplus)$ has a maximum at $x=\nplus$, with value
$f_{\mathrm{max}}=(\frac32-\ln 2)\nplus$.
Therefore, the integrand is uniformly bounded by a (decreasing) function of $n$
times $\alpha \esp{-\pi\alpha}$. Therefore also in this intermediate region the
matrix element is exponentially suppressed in $\alpha$.

\begin{figure}[ht!]
  \centering
  \includegraphics[width=0.6\linewidth]{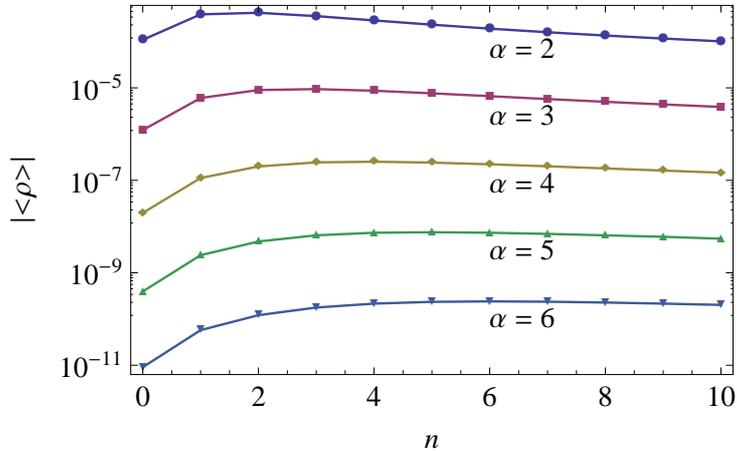}
  \caption{{\it Numerical computation of the matrix element~(\ref{bsme}) for $n$ up
    to 10. Each line represents a different value of $\alpha$, here ranging from
    2 to 6.}}
  \label{f:bsme}
\end{figure}

These results are confirmed by numerically computing the matrix
element~(\ref{bsme}), as shown in fig.~\ref{f:bsme}.  In conclusion, at least
in this first-order perturbative treatment, the $S$-matrix
eigenvalue~(\ref{pertRes}) does not receive significative enhancements from the
bound-states eigenfunctions beyond the exponentially suppressed contribution
$\psi_c(0)$.

%======================================================================
\subsection{Transitions to continuum states\label{ss:tcs}}
%======================================================================

The second mechanism of possible restoration of unitarity that we consider is
the transition towards ``high-energy'' eigenstates $\bra{\phi_E}$ of the
continuum spectrum of $H_c$. The motivation is that, for $\rho\leq0$, we expect
the corresponding eigenfunctions $\phi_E(\rho)$ to be less suppressed than
$\psi_c$, which is just the zero-energy eigenfunction. The eigenfunctions obey
the Schr\"odinger equation ($\hbar=1$)
\begin{equation}\label{Eeq}
  H_c\,\phi_E \equiv \left[-\frac{1}{4\alpha}\frac{\dif^2}{\dif\rho^2}
  +\alpha\left(\frac{1}{\rho}-1\right)\right]\phi_E = E\,\phi_E
\end{equation}
and the condition of having only transmitted wave for $\rho<0$. If we write the
energy $E=\alpha(\tz^2-1)$ in terms of the parameter $\tz\geq0$, it is easy to
check that, with the rescalings $\alpha\to\alpha/\tz$ and $\rho\to\rho\tz^2$,
eq.~(\ref{Eeq}) reduces to eq.~(\ref{schrodinger}) for $\psi_c=\phi_0$.
Therefore, the generic eigenfunction can be expressed in terms of $\psi_c$,
namely
\begin{equation}\label{phiE}
  \phi_E(\rho;\alpha) = \sqrt{\frac{\alpha}{\pi}} \;
  \psi_c\big(\tz^2\rho;\frac{\alpha}{t_0}\big) \;,
\end{equation}
where the prefactor $\sqrt{\alpha/\pi}$ has been chosen in order to obtain the
continuum normalization
\begin{equation}\label{contNorm}
  \int \phi_E^*(\rho) \, \phi_{E'}(\rho)\;\dif\rho = \delta(\tz-\tz') \;.
\end{equation}
Eq.~(\ref{phiE}) clearly shows that, for large energies $E\gg\alpha\iff\tz\gg1$,
the wave function at the origin
$\phi_E(0)\sim\esp{-\frac{\pi\alpha}{\tz}}$ is really
much larger than $\psi_c(0)$, because it is easier to cross the barrier at
$\rho\geq 0$.

It remains to evaluate the matrix elements $\bk{\phi_E|\rho|\psi_c}$. By using
the integral representation of $\phi_c$ derived in ref.~\cite{CC09} and the
relation~(\ref{phiE}), we can write
\begin{align}
  \phi_E(\rho) &= \sqrt{\frac{\alpha}{\pi}} N\big(\frac{\alpha}{\tz}\big)
  \int_{-\infty-\ui0}^{+\infty}
  \sign(t-1)\left(\frac{t-1}{t+1}\right)^{\ui\frac{\alpha}{\tz}}
  \frac{\esp{\ui2\alpha\rho t \tz}}{t^2-1} \;\dif t
 \label{intRep} \\
  N(\alpha) &\equiv
  \frac{\esp{-\frac{\pi\alpha}{2}}}{\cosh(\pi\alpha)\Gamma(\ui\alpha)} \;.
 \label{N}
\end{align}
With this representation we can explicitly perform the $\rho$ integration in the
matrix element, as outlined in app.~\ref{a:mecs}. The result is expressed in
terms of hypergeometric functions:
\begin{align}
  &\int_{-\infty}^{+\infty} \phi_E^*(\rho) \rho \psi_c(\rho) \;\dif\rho
  = N(\alpha) N^*\big(\frac{\alpha}{\tz}\big) \frac{\pi}{4\alpha^2\az}
  \frac{\partial^2}{\partial \tz^2} I(\tz)
 \label{contMatEl} \\
  &I(\tz) = \left(\frac{\tz-1}{\tz+1}\right)^{\ui(\alpha+\az)}
  \esp{\pi\az} \Big[\pi \az \zeta \coth(\pi\alpha) F(1-\ui\alpha,1-\ui\az;2;\zeta)
 \nonumber \\
  &\qquad\qquad\qquad +\frac{\Gamma(-\ui\alpha)\Gamma(1-\ui\az)}{\Gamma\big(1-\ui(\alpha+\az)\big)}
  F\big(-\ui\alpha,-\ui\az;1-\ui(\alpha+\az);1-\zeta\big) \Big] \;,
 \label{d:I}
\end{align}
where $\zeta\equiv -\frac{4\tz}{(1-\tz)^2}$ and $\az\equiv\frac{\alpha}{\tz}$.

Let us now study the order of magnitude of $I(\tz)$ which determines the matrix
element to continuum states. We firstly consider transitions to high-energy
states ($\tz^2=1+\frac{E}{\alpha}\gg1$). In this case, the variable $\zeta\simeq-\frac{4}{\tz}$ in
$I(\tz)$ tends to zero, so that the two hypergeometric functions assume finite
values and cannot give exponentially enhanced contributions. Therefore,
the order of magnitude of $I(\tz)$ is determined by the factor $\esp{\pi\az}$ in
eq.~(\ref{d:I}). By taking into account the wave-function normalizations, the
matrix element is of order
\begin{equation}\label{matrElOrd}
  N(\alpha)N^*\big(\frac{\alpha}{\tz}\big)I(\tz)\sim
  \frac{\esp{-\frac{\pi\alpha}{2}}\esp{-\frac{\pi\az}{2}}}{\cosh(\pi\alpha)
   \cosh(\pi\az)\Gamma(\ui\alpha)\Gamma(\ui\az)}\esp{\pi\az}
  \sim\esp{-\pi\alpha}
\end{equation}
This result shows that the matrix element for transitions to continuum states of
very high energy suffers a suppression comparable to that of tunneling. As a
consequence, unitarity cannot be restored by these quantum effects.

Secondly, we consider transitions to lower energy states characterized by
$\tz\sim1$ but $\alpha-\az=\frac{\alpha}{\tz}(\tz-1)\gg1$, corresponding to a
quite large jump in energy, implying a lower suppression of
$\phi_E(0)\sim\esp{-\pi\az}$. In this regime, $\zeta\to-\infty$ and the
corresponding behaviour of the hypergeometric function is
\begin{align*}
  F(1-\ui\alpha,1-\ui\az;2;\zeta) &\sim \esp{\pi\az} \\
  F\big(-\ui\alpha,-\ui\az;1-\ui(\alpha+\az);1-\zeta\big) &\sim
  A_1 \esp{-\pi(\alpha-\az)}\esp{-\pi\alpha}+A_2\esp{-\pi\az} \;.
\end{align*}
The leading term is provided by the first hypergeometric function, yielding
$I(\tz)\sim\esp{2\pi\az}$. By inserting the wave-function normalization factors
(of order $\esp{-\pi(\alpha+\az)}$), the tunneling amplitude
\begin{equation}\label{tunAmpLe}
  \phi_E(0) \bk{\phi_E|\rho|\psi_c} \sim \esp{-\pi\az} \esp{-\pi(\alpha-\az)}
  \sim \esp{-\pi\alpha}
\end{equation}
turns out to be exponentially suppressed also in this case.

In conclusion, the first-order perturbative contribution to the
eigenvalue~(\ref{pertRes}) are exponentially suppressed by the matrix element in
all those cases where we expected sizeable effects thanks to the enhancement of
the eigenfuncions at the origin. Therefore, the quantum effects do not modify
the semiclassical picture described in sec.~\ref{s:sud}.

%%%%%%%%%%%%%%%%%%%%%%%%%%%%%%%%%%%%%%%%%%%%%%%%%%%%%%%%%%%%%%%%%%%%%%%%%%%%%%%%
\section{$\bs{\rho(0)}$-fluctuations and short-distance singularities\label{s:rhof}}
%%%%%%%%%%%%%%%%%%%%%%%%%%%%%%%%%%%%%%%%%%%%%%%%%%%%%%%%%%%%%%%%%%%%%%%%%%%%%%%%

In the preceding sections we have definitely confirmed the unitarity defect
previously found at semiclassical level for $b<b_c$~\cite{CC09}: indeed, the
critical behaviour of the action on the physical solutions around $b\simeq b_c$
is as expected (sec.~\ref{s:sud}) and the extra-contributions to the $S$-matrix
elements due to quantum transitions are likewise suppressed (sec.~\ref{s:sqt}).
However, we have kept throughout the analysis the ACV boundary conditions
$\dot\rho(\infty)=1$ and $\rho(0)=0$. While the former is needed in order to
match the large-distance behaviour with perturbative gravity and is thus
unavoidable, the latter insures that the solutions being considered are UV-safe,
so as to make the effective action self-sufficient. Is this really required?
What happens if we let $\rho(0)$ fluctuate?

Let us recall that a non-vanishing value of $\rho(0)\equiv\rho_0$ (of order $R^2$)
implies singularities at $r^2=0$ of $\dot\phi(r^2)$ and of $h(r^2)$, as follows
\begin{equation}\label{sing}
  (2\pi)^2 \dot\phi \simeq -\frac{\rho_0}{r^2} \;, \qquad
  h = \nabla^2\phi \simeq -\frac{\rho_0}{\pi^2}\delta(r^2) \;,
\end{equation}
where the latter (implied by the outgoing flux of $\nabla\phi$) can be
interpreted by assigning $\rho(r^2)$ a discontinuity $\rho_0$ at $r^2=0$ so as to
have
\begin{equation}\label{hDecomp}
  (\pi R)^2 h(r^2) = (1-\dot\rho)_\reg -\rho_0 R^2 \delta(r^2) \;.
\end{equation}
Although the form of such singularities is probably model dependent, their
existence is expected for generic boundary conditions.
They stem from the very fact that $R^2$ is the dimensionful coupling of the
2-dimensional action~(\ref{2dimAction}) and in fact they occurred already in the
more general 2-dimensional treatment of~\cite{ACV93}.

The above singularities produce a singular metric~(\ref{metr}) and a divergent
action. If we take them seriously, the divergent part of the action in
eq.~(\ref{d:Sq}) is
\begin{align}
  \A &= -\alpha\int_0^\infty \left[(1-\dot\rho)^2 -
    \frac{R^2}{\rho}\Theta(r^2-b^2)\right] \; \frac{\dif r^2}{R^2}
  + \sqrt{2\alpha y} \int_0^\infty (1-\dot\rho)[a(r^2)+a^\dagger(r^2)] \;
  \frac{\dif r^2}{R}
 \label{divAct} \\
  &= \A_\div + \A_\reg
  \;, \qquad\qquad
  \A_\div(\rho_0) \simeq -\alpha\rho_0^2 \delta(0)
  - \sqrt{2\alpha y \rho_0 \delta(0)} \, (a_0+a_0^\dagger) \;,
 \label{Adiv}
\end{align}
where the formally divergent $\delta(0)\simeq\ord{\frac{R^2}{\lambda_s^2}}$ is
presumably regularized by the string in the ACV approach. Note that the
divergence affects not only the c-number part of $\A$, but also the $r^2=0$ mode
of the coherent state operator, that we have normalized to $
[a_0,a_0^\dagger]=1$.

At this point we notice that the divergent part of the $S$-matrix suppresses in a
drastic way elastic and quasi-elastic processes with $\rho_0\neq 0$ because
\begin{equation}\label{qesupp}
  \bk{0|\esp{\ui\A_\div(\rho_0)}|0} \sim \esp{-\alpha(\ui+y)\rho_0^2\delta(0)}
  \simeq \esp{-\alpha(\ui+y)\rho_0^2 \frac{R^2}{\lambda_s^2}}
\end{equation}
shows violent oscillations and absorption, with exponent of order $\sim
\alpha\rho_0^2\frac{R^2}{\lambda_P^2}\sim\alpha^2\rho_0^2$ for
$\lambda_s\sim\lambda_P=\sqrt{G\hbar}$. Therefore, at quasi-elastic level, we
were justified in setting $\rho_0=0$ in the first place. And that would close
the argument.

The above conclusion is not fully satisfactory, though. On one hand,
$\rho_0\neq0$ is associated to a singular metric that we think corresponds to
classically trapped solutions~\cite{EdGi02,YoNa03,GiRy04,KoVe02,VeWo08}, and we
would like to know about their fate at quantum level. On the other hand, if we
think of the action~(\ref{divAct}) from the unitarity point of view, the
divergent part is, after all, a hermitian operator and we expect it to
contribute a unitary $S$-matrix if we sum over all possible emissions. This
statement is confirmed by the inclusive equations~(sec.~\ref{s:sud}) because we
do have real-valued solutions for $b<b_c$ also, provided we take
$\rho_0=\tilde\rho_0=\rho_m(b)>0$, as mentioned at the end of
sec.~\ref{s:bccip}. In this cases, the inclusive action vanishes: should we
conclude that considering $\rho_0\neq0$ insures $S$-matrix unitarity of the
model?

The trouble with our second argument is that it is inconsistent with energy
conservation. In fact, the action~(\ref{divAct}) provides total probabilities of
order unity only in association with quite a number
$\sim\alpha\frac{R^2}{\lambda_s^2}$ of hard emitted gravitons with energies of
the order of the Planck mass. Barring other dynamical effects (possibly
providing a red-shift), this would require an energy which is much larger than
$\sqrt{s}$.

Nevertheless, from the above considerations, our tentative conclusion is that,
from the unitarity stand point, we {\em should} actually consider the solutions
with $\rho_0\neq0$ also by keeping in mind that the present model is inadequate
in the short-distance region and we should therefore come back to the string
dynamics, that was originally neglected in the regime $R\gg\lambda_s$. The role
of the latter is not only that of regularizing the would-be singularities at the
string scale, but also of providing a number of physical effects at distances
which are intermediate between $\lambda_s$ and $R$.

For instance, a well-known~\cite{ACV88,GiGrMa07} effect, occurring already at
distances of order $R$, is string excitation by tidal forces (or ``diffractive''
string excitation). It looks straightforward to estimate it in the present
model, but it has not been done yet. Another issue, perhaps more important for
the unitarity problem, is the evaluation of rescattering
corrections~\cite{ACV93}.  Here the string is needed in order to regulate the UV
divergencies in the longitudinal variables so as to extract a finite answer. We
expect from them an improvement of the longitudinal dynamics%
\footnote{For instance, we expect the wave-fronts of the
  solution~(\ref{metrica}) to be shifted~\cite{tH87} when they merge each other
  just to make their motion consistent with the scattering process.}
for $b=\ord{R}$ and perhaps some hint as to the existence of trapped solutions
for $\rho_0\neq0$. If effects of this kind turn out to yield an important
contribution to unitarity, then this will involve relatively soft gravitons, of
energies $\sim\hbar/R$ and, for this reason, may be consistent with energy
conservation.

It may be also that string effects are mostly confined at scale $\lambda_s$. In
such cases we expect that compactified dimensions and/or universes can be
excited and, therefore, that probability --- in our original scattering process
--- is lost for good reasons: in order to recover it  we would need scattering
data in some extra world.%
\footnote{An example of this kind is perhaps provided by the recent analysis of
  string-brane scattering~\cite{DDRV10} in the case that the metric potential is
  singular enough to allow fall into the center.}
Whichever the case, we are inclined to think that, for $b<b_c$, solutions with
$\rho(0)\neq0$, depending on the string dynamics, can play an important role in
explaining the unitarity loss of the present model.

%%%%%%%%%%%%%%%%%%%%%%%%%%%%%%%%%%%%%%%%%%%%%%%%%%%%%%%%%%%%%%%%%%%%%%%%%%%%%%%%
\section{Discussion\label{s:d}}
%%%%%%%%%%%%%%%%%%%%%%%%%%%%%%%%%%%%%%%%%%%%%%%%%%%%%%%%%%%%%%%%%%%%%%%%%%%%%%%%

The first and most detailed outcome of this paper is that the reduced-action
model for the transplanckian $S$-matrix, valid in the regime $R,b\gg\lambda_s$,
shows a unitarity suppression in the region $b<b_c$, that we think corresponds
to classical gravitational collapse. This feature, originally found at
semiclassical level in~\cite{CC09}, is here confirmed at more general quantum
level in sec.~\ref{s:sqt} and by a perturbative method around the critical
region in sec.~\ref{s:sud}. We find in particular that the unitarity defect
around the critical point is non analytic with a fractional critical exponent
when inelastic emission is forbidden ($y=0$), while it shows a more regular
behaviour when $y>0$.

Since such unitarity defect raises questions about the information loss in
gravitational collapse, we have tried to investigate also whether the present
model is complete and if not, how to complete it. Let us first note that the
model has a hermitian lagrangian and is thus expected to be unitary, barring
some special conditions. But a key role is played by the boundary condition
$\rho(0)=0$. On one hand, it is needed in order to provide UV-safe solutions
with regular metric, so as to insure that the eikonal is a perturbative series
in $R^2/b^2$ for $b>b_c$, independently of the string scale $\lambda_s$. On
the other hand, that condition is the main cause of the unitarity defect for
$b<b_c$ because, in order to reach $\rho=0$ at $r^2=0$, the amplitude has to
cross a potential barrier in $\rho$-space by a tunnel effect, and is thereby
suppressed.

For the above reasons, here we suggest to consider the solutions with
$\rho(0)\neq 0$ as the alternative path that the probability flow may take. Such
solutions yield a singular metric and a divergent action and probably correspond
to classically trapped fields for $b<b_c$. In the toy model of sec.~\ref{s:rhof}
we let $\rho(0)$ fluctuate with a weight provided by the reduced action itself,
and we find that $\rho(0)\neq0$ is violently suppressed at elastic and
quasi-elastic level, thus justifying the condition $\rho(0)=0$ in those
cases. Nevertheless, the model is formally unitary and acquires a total
probability of order unity in association with a large number of hard
gravitons, a process which is, however, forbidden by energy conservation.

The paradoxical features above illustrate the problem we have in looking at the
$\rho(0)\neq0$ solutions: the present model suggests that unitarity might be
recovered at inelastic level, but is by itself inadequate. Therefore, we have to
rely on the string dynamics and related vertices~\cite{ABC89} in order to
describe the short-distance behaviour of the solutions and their contributions
to unitarity. A key point is to determine the important scales of the evolution
from $R$ to $\lambda_s$, and the kind of inelastic channels which are involved,
which may include states propagating in extra dimensions or universes. In
sec.~\ref{s:rhof} we have listed several string effects, some of which at scale
$\lambda_s$ and some at scales of order $R$, but only a careful analysis can
tell us where the lost probability goes.

%%%%%%%%%%%%%%%%%%%%%%%%%%%%%%%%%%%%%%%%%%%%%%%%%%%%%%%%%%%%%%%%%%%%%%%%%%%%%%%%
\section*{Acknowledgements}
%%%%%%%%%%%%%%%%%%%%%%%%%%%%%%%%%%%%%%%%%%%%%%%%%%%%%%%%%%%%%%%%%%%%%%%%%%%%%%%%

It is a pleasure to thank Giuseppe D'Appollonio, Paolo Di Vecchia, Rodolfo Russo
and Gabriele Veneziano for stimulating discussions, during the workshop
``Large-$N$ Gauge Theories'' at the Galileo Galilei Institute, Florence, April
2011.

%%%%%%%%%%%%%%%%%%%%%%%%%%%%%%%%%%%%%%%%%%%%%%%%%%%%%%%%%%%%%%%%%%%%%%%%%%%%%%%%%
%%%%%%%%%%%%%%%%%%%%%%%%%%%%%%%%%%%%%%%%%%%%%%%%%%%%%%%%%%%%%%%%%%%%%%%%%%%%%%%%%
\appendix
%%%%%%%%%%%%%%%%%%%%%%%%%%%%%%%%%%%%%%%%%%%%%%%%%%%%%%%%%%%%%%%%%%%%%%%%%%%%%%%%%
%%%%%%%%%%%%%%%%%%%%%%%%%%%%%%%%%%%%%%%%%%%%%%%%%%%%%%%%%%%%%%%%%%%%%%%%%%%%%%%%%

%%%%%%%%%%%%%%%%%%%%%%%%%%%%%%%%%%%%%%%%%%%%%%%%%%%%%%%%%%%%%%%%%%%%%%%%%%%%%%%%%
\section{Matrix element to continuum states\label{a:mecs}}
%%%%%%%%%%%%%%%%%%%%%%%%%%%%%%%%%%%%%%%%%%%%%%%%%%%%%%%%%%%%%%%%%%%%%%%%%%%%%%%%%

In this appendix we explicitly perform the computation of the matrix element to
continuum states of eq.~(\ref{contMatEl}). Starting from the integral
representation~(\ref{intRep}) for $\phi_E$ and $\psi_c=\phi_0$ we have
\begin{align}
  \int_{-\infty}^{+\infty}\phi_E^*\,\rho\,\psi_c \;\dif\rho
  &=N(\alpha)N^*\big(\frac{\alpha}{\tz}\big)\int_{-\infty}^{+\infty}
  \dif t\, \dif p\, \dif\rho
 \nonumber \\
  &\quad \times \rho\,\esp{\ui2\alpha\rho(t-\tz p)}
  \left(\frac{t-1}{t+1}\right)^{\ui\alpha}
  \left(\frac{p-1}{p+1}\right)^{-\ui\frac{\alpha}{\tz}}
  \frac{\sign(t-1)\sign(p-1)}{(t^2-1)(p^2-1)} \;.
 \label{riem}
\end{align}
The integration in $\rho$ can be performed by introducing
$\az\equiv\frac{\alpha}{\tz}$ and rewriting the integrand as
\begin{equation}\label{newintegrand}
  \frac{\ui}{2\alpha p}\partial_{\tz} \esp{\ui 2\alpha\rho(t-\tz p)}
  \left(\frac{t-1}{t+1}\right)^{\ui\alpha}
  \left(\frac{p-1}{p+1}\right)^{-\ui\az}
  \frac{\sign(t-1)\sign(p-1)}{(t^2-1)(p^2-1)} \;.
\end{equation}
It produces the delta function $\delta\big(2\alpha(t-\tz p)\big)$ which is used
to perform the integration in $t$. Apart from the $N$ normalization factors, the
matrix element is now
\begin{equation}\label{deint1}
  \frac{\ui\pi}{2\alpha^2}\partial_{\tz}\int
  \left(\frac{\tz p-1}{\tz p+1}\right)^{\ui\alpha}
  \left(\frac{p-1}{p+1}\right)^{-\ui\az}
  \frac{\sign(\tz p-1)\sign(p-1)}{(\tz^2 p^2-1)(p^2-1)} \;\frac{\dif p}{p} \;.
\end{equation}
By rescaling the variable $\tz p \to p$ the four branch points are found at
$p=\pm\tz,\pm1$ and the integrand gets a factor $t_0^2$ that we conveniently
rewrite as $[\tz^2-p^2]+p^2$, obtaining
\begin{align}
  \text{(\ref{deint1})} &= -\frac{\ui\pi}{2\alpha^2}\partial_{\tz}\int
  \left(\frac{p-1}{p+1}\right)^{\ui\alpha}
  \left(\frac{p-\tz}{p+\tz}\right)^{-\ui\az}
  \frac{\sign(p-1)\sign(p-\tz)}{p^2-1} \;\frac{\dif p}{p}
 \nonumber \\
 &\quad +\frac{\ui\pi}{2\alpha^2}\partial_{\tz}\int
  \left(\frac{p-1}{p+1}\right)^{\ui\alpha}
  \left(\frac{p-\tz}{p+\tz}\right)^{-\ui\az}
  \frac{\sign(p-1)\sign(p-\tz)}{(p^2-1)(p^2-\tz^2)} p \;\dif p \;.
\end{align}
Let us now consider the two integrals separately.  If we perform the $\tz$
derivative in the first integral, we obtain
\begin{equation}\label{firstInt}
  \frac{\ui\pi}{2\alpha^2}\int
  \left(\frac{p-1}{p+1}\right)^{\ui\alpha}
  \left(\frac{p-\tz}{p+\tz}\right)^{-\ui\az}
  \frac{\sign(p-1)\sign(p-\tz)}{(p^2-1)(p^2-\tz^2)} \;\dif p
\end{equation}
which vanishes, being proportional to the scalar product $\bk{\phi_E|\phi_0}$ of
two eigenfunctions with different energy. Only the second integral remains, and
it can be simplified by noting that the factor $p/(p^2-t_0^2)$ can be obtained
by a further derivative with respect to $\tz$. In practice, 
\begin{align}
  \text{(\ref{deint1})} &= \frac{\pi}{4\alpha^2\az}\partial_{\tz}^2 I(\tz)
 \nonumber \\
 I(\tz) &\equiv \int
  \left(\frac{p-1}{p+1}\right)^{\ui\alpha}
  \left(\frac{p-\tz}{p+\tz}\right)^{-\ui\az}
  \frac{\sign(p-1)\sign(p-\tz)}{p^2-1} \;\dif p \;.
\end{align}
The integrand of $I(\tz)$ is built by four powers whose exponents' sum is
$-2$. This allows us to express it in terms of the hypergeometric function, as
follows. By a linear change of integration variable
$x\equiv\frac{1-\tz}{2}\frac{p+1}{p-\tz}$, we can map three of the four branch points
into the ``canonical'' ones 0, 1, $\infty$:
\begin{equation}\label{01inf}
  I(\tz)=\frac12\left(\frac{\tz-1}{\tz+1}\right)^{\ui(\alpha+\az)}
  \int_{-\infty}^{+\infty} \sign(x-1) \left(\frac{x-1}{x}\right)^{\ui\alpha}
  [-(1-\zeta x)]^{\ui\az} \frac{\dif x}{x(1-x)}
\end{equation}
where $\zeta\equiv\frac{-4\tz}{(1-\tz)^2}$. Here the integration path is
slightly shifted off the real axis so as to lie below the cut $[0,1]$ and above
the cut from $-1/\zeta$ to infinity. The sign in the integrand amounts to split
the integration into two pieces: one from $-\infty$ to $1$ and one
from $+\infty$ to $1$. By rotating the latter around 1 in counterclockwise
direction, both pieces ranges from $-\infty$ to $1$, one below and one above the
$[0,1]$ cut. In this way, we arrive at the expression
\begin{align}
  I(\tz) = \frac12\left(\frac{\tz-1}{\tz+1}\right)^{\ui(\alpha+\az)}
  \esp{\pi\az}& \left[- 2\cosh(\pi\alpha) \int_0^1
  x^{-\ui\alpha-1}(1-x)^{\ui\alpha-1}(1-\zeta x)^{\ui\az} \;\dif x \right.
 \nonumber \\
  &\left.\;+2\int_0^\infty x^{-\ui\alpha-1}(1+x)^{\ui\alpha-1}
  (1+\zeta x)^{\ui\az} \;\dif x \right] \;,
 \label{twoint}
\end{align}
where the $2\cosh(\pi\alpha)$ factor arises from evaluating the integrand above
and below the $[0,1]$ cut.  With the substitution $y=\frac{x}{1+x}$ in the
second integral, we recognize in eq.~(\ref{twoint}) two integral representations
of the hypergeometric function, yielding
\begin{align}
  I(\tz) = \left(\frac{\tz-1}{\tz+1}\right)^{\ui(\alpha+\az)}
 & \esp{\pi\az}  \Big[ \pi\az\zeta\coth(\pi\alpha)
  F(1-\ui\alpha,1-\ui\az;2;\zeta)
 \nonumber \\
  &+\frac{\Gamma(-\ui\alpha)\Gamma(1-\ui\az)}{\Gamma\big(1-\ui(\alpha+\az)\big)}
  F\big(-\ui\alpha,-\ui\az;1-\ui(\alpha+\az);1-\zeta\big)\Big]
 \label{Itz}
\end{align}
as in eq.~(\ref{d:I}).

%%%%%%%%%%%%%%%%%%%%%%%%%%%%%%%%%%%%%%%%%%%%%%%%%%%%%%%%%%%%%%%%%%%%%%%%%%%%%%%%


\begin{thebibliography}{99}
%%%%%%%%%%%%%%%%%%%%%%%%%%%%%%%%%%%%%%%%%%%%%%%%%%%%%%%%%%%%%%%%%%%%%%%%%%%%%%%%

\bibitem{ACV88}
  D.~Amati, M.~Ciafaloni and G.~Veneziano,
  ``Superstring Collisions at Planckian Energies,''
  Phys.\ Lett.\  B {\bf 197} (1987) 81;
  %%CITATION = PHLTA,B197,81;%%
  ``Classical and quantum gravity effects from planckian energy superstring
  collisions,''
  Int.\ J.\ Mod.\ Phys.\  A {\bf 3} (1988) 1615;
  %%CITATION = IMPAE,A3,1615;%%
  ``Can Space-Time Be Probed Below The String Size?,''
  Phys.\ Lett.\  B {\bf 216} (1989) 41.
  %%CITATION = PHLTA,B216,41;%%

\bibitem{GrMe87}
  D.~J.~Gross and P.~F.~Mende,
  ``The High-Energy Behavior of String Scattering Amplitudes,''
  Phys.\ Lett.\  B {\bf 197} (1987) 129;
  %%CITATION = PHLTA,B197,129;%%
  ``String Theory Beyond the Planck Scale,''
  Nucl.\ Phys.\  B {\bf 303} (1988) 407.
  %%CITATION = NUPHA,B303,407;%%

\bibitem{ACV90}
  D.~Amati, M.~Ciafaloni and G.~Veneziano,
  ``Higher Order Gravitational Deflection And Soft Bremsstrahlung In Planckian
  Energy Superstring Collisions,''
  Nucl.\ Phys.\  B {\bf 347} (1990) 550.
  %%CITATION = NUPHA,B347,550;%%

\bibitem{ACV93}
  D.~Amati, M.~Ciafaloni and G.~Veneziano,
  ``Effective action and all order gravitational eikonal at Planckian
  energies,''
  Nucl.\ Phys.\  B {\bf 403}, 707 (1993).
  %%CITATION = NUPHA,B403,707;%%

\bibitem{ACV07}
  D.~Amati, M.~Ciafaloni and G.~Veneziano,
  ``Towards an $S$-matrix Description of Gravitational Collapse,''
  JHEP {\bf 0802} (2008) 049
  [arXiv:0712.1209 [hep-th]].
  %%CITATION = JHEPA,0802,049;%%

\bibitem{CC08}
  M.~Ciafaloni and D.~Colferai,
  ``S-matrix and Quantum Tunneling in Gravitational Collapse,''
  JHEP {\bf 0811} (2008) 047
  [arXiv:0807.2117 [hep-th]].
  %%CITATION = JHEPA,0811,047;%%

\bibitem{CC09}
  M.~Ciafaloni and D.~Colferai,
  ``Quantum Tunneling and Unitarity Features of an $S$-matrix for Gravitational
  Collapse,''
  JHEP {\bf 0912} (2009) 062
  [arXiv:0909.4523 [hep-th]].
  %%CITATION = JHEPA,0912,062;%%

\bibitem{Hawking}
  S.~W.~Hawking,
  ``Black hole explosions,''
  Nature {\bf 248} (1974) 30;
  %%CITATION = NATUA,248,30;%%
  ``Particle Creation By Black Holes,''
  Commun.\ Math.\ Phys.\  {\bf 43} (1975) 199
  [Erratum-ibid.\  {\bf 46} (1976) 206];
  %%CITATION = CMPHA,43,199;%%
  ``Black holes and the information paradox,''
  %\href{/spires/find/hep/www?irn=6536760}{SPIRES entry}
  {\it Prepared for GR17: 17th International Conference on General Relativity
    and Gravitation, Dublin, Ireland, 18-24 Jul 2004};
  ``Information Loss in Black Holes,''
  Phys.\ Rev.\  D {\bf 72} (2005) 084013
  [arXiv:hep-th/0507171].
  %%CITATION = PHRVA,D72,084013;%%

\bibitem{Li82}
  L.~N.~Lipatov,
  ``Multi-Regge processes in gravitation,''
  Sov.\ Phys.\ JETP {\bf 55} (1982) 582
  [Zh.\ Eksp.\ Teor.\ Fiz.\  {\bf 82} (1982) 991].
  %%CITATION = ZETFA,82,991;%%

\bibitem{ABC89}
  M.~Ademollo, A.~Bellini and M.~Ciafaloni,
  ``Superstring Regge amplitudes and emission vertices,''
  Phys.\ Lett.\  B {\bf 223} (1989) 318;
  %%CITATION = PHLTA,B223,318;%%
  ``Superstring Regge amplitudes and graviton radiation at planckian energies,''
  Nucl.\ Phys.\  B {\bf 338} (1990) 114.
  %%CITATION = NUPHA,B338,114;%%

\bibitem{Li91}
  L.~N.~Lipatov,
  ``High-energy scattering in QCD and in quantum gravity and two-dimensional
  field theories,''
  Nucl.\ Phys.\  B {\bf 365} (1991) 614.
  %%CITATION = NUPHA,B365,614;%%

\bibitem{KiSz95}
  R.~Kirschner and L.~Szymanowski,
  ``Effective action for high-energy scattering in gravity,''
  Phys.\ Rev.\  D {\bf 52} (1995) 2333
  [arXiv:hep-th/9412087].
  %%CITATION = PHRVA,D52,2333;%%

\bibitem{Ve93}
  E.~P.~Verlinde and H.~L.~Verlinde,
  ``High-energy scattering in quantum gravity,''
  Class.\ Quant.\ Grav.\  {\bf 10} (1993) S175.
  %%CITATION = CQGRD,10,S175;%%

\bibitem{CVprep}
  M.~Ciafaloni and G.~Veneziano, unpublished.

\bibitem{AGK73}
  V.~A.~Abramovsky, V.~N.~Gribov and O.~V.~Kancheli,
  ``Character of inclusive spectra and fluctuations produced in inelastic
  processes by multi-pomeron exchange,''
  Yad.\ Fiz.\  {\bf 18} (1973) 595
  [Sov.\ J.\ Nucl.\ Phys.\  {\bf 18} (1974) 308].
  %%CITATION = SJNCA,18,308;%%

\bibitem{AS}
  see, e.g., M.~Abramowitz and I.A.~Stegun,
  ``Handbook of mathematical functions'', Dover publications.

\bibitem{EdGi02}
  D.~M.~Eardley and S.~B.~Giddings,
  ``Classical black hole production in high-energy collisions,''
  Phys.\ Rev.\  D {\bf 66} (2002) 044011
  [arXiv:gr-qc/0201034].
  %%CITATION = PHRVA,D66,044011;%%

\bibitem{YoNa03}
  H.~Yoshino and Y.~Nambu,
  ``Black hole formation in the grazing collision of high-energy particles,''
  Phys.\ Rev.\  D {\bf 67} (2003) 024009
  [arXiv:gr-qc/0209003].
  %%CITATION = PHRVA,D67,024009;%%

\bibitem{GiRy04}
  S.~B.~Giddings and V.~S.~Rychkov,
  ``Black holes from colliding wavepackets,''
  Phys.\ Rev.\  D {\bf 70} (2004) 104026
  [arXiv:hep-th/0409131].
  %%CITATION = PHRVA,D70,104026;%%

\bibitem{KoVe02}
  E.~Kohlprath and G.~Veneziano,
  ``Black holes from high-energy beam-beam collisions,''
  JHEP {\bf 0206} (2002) 057
  [arXiv:gr-qc/0203093].
  %%CITATION = JHEPA,0206,057;%%

\bibitem{VeWo08}
  G.~Veneziano and J.~Wosiek,
  ``Exploring an $S$-matrix for gravitational collapse,''
  arXiv:0804.3321 [hep-th];
  %%CITATION = ARXIV:0804.3321;%%
  ``Exploring an $S$-matrix for gravitational collapse II: a momentum space
  analysis,''
  arXiv:0805.2973 [hep-th].\\
  %%CITATION = ARXIV:0805.2973;%%
  G.~Marchesini and E.~Onofri,
  %``High energy gravitational scattering: a numerical study,''
  JHEP {\bf 0806}, 104 (2008)
  [arXiv:0803.0250 [hep-th]].
  %%CITATION = JHEPA,0806,104;%%

\bibitem{GiGrMa07}
  S.~B.~Giddings, D.~J.~Gross and A.~Maharana,
  ``Gravitational effects in ultrahigh-energy string scattering,''
  Phys.\ Rev.\  D {\bf 77} (2008) 046001
  [arXiv:0705.1816 [hep-th]].
  %%CITATION = PHRVA,D77,046001;%%

\bibitem{tH87}
  G.~'t Hooft,
  ``Graviton Dominance in Ultrahigh-Energy Scattering,''
  Phys.\ Lett.\  B {\bf 198} (1987) 61.
  %%CITATION = PHLTA,B198,61;%%

\bibitem{DDRV10}
  G.~D'Appollonio, P.~Di Vecchia, R.~Russo and G.~Veneziano,
  ``High-energy string-brane scattering: leading eikonal and beyond,''
  JHEP {\bf 1011} (2010) 100
  [arXiv:1008.4773 [hep-th]].
  %%CITATION = JHEPA,1011,100;%%

\end{thebibliography}
\end{document}